\documentclass[acmtocl]{acmtrans2m}

\acmVolume{?}
\acmNumber{?}
\acmYear{??}
\acmMonth{?}

\newtheorem{theorem}{Theorem}[section]
\newtheorem{corollary}[theorem]{Corollary}
\newtheorem{lemma}[theorem]{Lemma}
\newdef{definition}[theorem]{Definition}
\newdef{remark}[theorem]{Remark}
\newdef{example}[theorem]{Example}

\usepackage{latexsym}
\usepackage{amsfonts}

\newbox\tempa
\newbox\tempb
\newdimen\tempc
\newbox\tempd

\def\mud#1{\hfil $\displaystyle{#1}$\hfil}
\def\rig#1{\hfil $\displaystyle{#1}$}

\def\inruleanhelp#1#2#3{\setbox\tempa=\hbox{$\displaystyle{\mathstrut #2}$}%
                        \setbox\tempd=\hbox{$\; #3$}%
                        \setbox\tempb=\vbox{\halign{##\cr
        \mud{#1}\cr
        \noalign{\vskip\the\lineskip}%
        \noalign{\hrule height 0pt}%
        \rig{\vbox to 0pt{\vss\hbox to 0pt{\copy\tempd \hss}\vss}}\cr
        \noalign{\hrule}%
        \noalign{\vskip\the\lineskip}%
        \mud{\copy\tempa}\cr}}%
                      \tempc=\wd\tempb
                      \advance\tempc by \wd\tempa
                      \divide\tempc by 2 }

\def\inrulean#1#2#3{{\inruleanhelp{#1}{#2}{#3}%
                     \hbox to \wd\tempa{\hss \box\tempb \hss}}}

\def\ian#1#2#3{{\lineskip 4pt\inrulean{#1}{#2}{#3}}}

\def\lowerhalf#1{\hbox{\raise -0.8\baselineskip\hbox{#1}}}

\def\ianc#1#2#3{{\lineskip 4pt\lowerhalf{\inruleanhelp{#1}{#2}{#3}%
                   \box\tempb\hskip\wd\tempd}}}
\def\ibnc#1#2#3#4{{\lineskip 4pt\ianc{#1\quad\qquad #2}{#3}{#4}}}

\newcommand{\arv}[2]{\begin{array}{c} #1\\#2\end{array}}

\newcommand{\Imp}{\supset}

\newcommand{\And}{\wedge}
\newcommand{\BImp}{\Leftrightarrow}

\newcommand{\itlam}{{\it lam}}
\newcommand{\itapp}{{\it app}}
\newcommand{\itexp}{{\it exp}}

\newcommand{\lamb}[2]{\lambda {#1}\oftp{#2}\ldot}
\newcommand{\lxua}{\lam x^u\oftp A\ldot} %  ^u added: A.M.

\newcommand{\ldot}{.\;}

\newcommand{\bnfas}{\mathrel{::=}}
\newcommand{\bnfalt}{\mid}
\newcommand{\arrow}{\rightarrow}
\newcommand{\lam}{\lambda}
\newcommand{\oftp}{\mathord{:}}
\newcommand{\hastype}{\mathrel{:}}
\newcommand{\rep}[1]{\ulcorner #1\urcorner}  % amsfonts.sty
\newcommand{\vd}{\vdash}
\newcommand{\nvd}{\null\mathrel{\not\vdash}}

\newcommand{\one}{\mathbf{1}}
\newcommand{\zero}{\mathbf{0}}

\newcommand{\GG}{\Gamma}
\newcommand{\DD}{\Delta}
\newcommand{\OO}{\Omega}
\newcommand{\god}{\Gamma; \Omega; \Delta}
\newcommand{\godc}{(\GG,\OO,\DD)}

\newcommand{\cD}{{\cal D}}
\newcommand{\cE}{{\cal E}}
\newcommand{\cM}{{\cal M}}
\newcommand{\cN}{{\cal N}}
\newcommand{\cP}{{\cal P}}

\newcommand{\da}{\downarrow}
\newcommand{\bua}{\Uparrow}
\newcommand{\buda}{\Uparrow\!\downarrow}
\newcommand{\blip}{\Rightarrow}

\newcommand{\whr}{\stackrel{\mbox{\tiny whr}}{\longrightarrow}}
\newcommand{\betared}{\stackrel{\beta}{\longrightarrow}}

\newcommand{\etaexp}{\stackrel{\bar\eta}{\longrightarrow}}

\newcommand{\GImp}[1]{\stackrel{\mbox{\tiny  ${#1}$}}{\arrow}}
\newcommand{\TImp}{\GImp{u}}
\newcommand{\RImp}{\GImp{1}}
\newcommand{\ZImp}{\GImp{0}}

\newcommand{\lop}{\mathbin{-\!\circ}}

\newcommand{\dom}{\mbox{\textrm{dom}}}
\newcommand{\mnot}{{\rm Not}}
\newcommand{\Pat}{\mathrm{Pat}_A(\Psi)}

\newcommand{\lsemantics}{\mathopen{\lbrack\mkern-3mu\lbrack}}
\newcommand{\rsemantics}{\mathclose{\rbrack\mkern-3mu\rbrack}}
\newcommand{\den}[1]{\lsemantics#1\rsemantics}   

\title{Higher-Order Pattern Complement and the Strict $\lambda$-Calculus}

\author{ALBERTO MOMIGLIANO\\University of Leicester\and
FRANK PFENNING\\Carnegie Mellon University}

\begin{abstract}
 We address the problem of complementing higher-order patterns without
repetitions of existential variables.  Differently from the first-order case,
the complement of a pattern cannot, in general, be described by a
pattern, or even by a finite set of patterns.  We therefore generalize
the simply-typed $\lambda$-calculus to include an internal notion of
\emph{strict function} so that we can directly express that a term must
depend on a given variable.  We show that, in this more expressive
calculus, finite sets of patterns without repeated variables are closed
under complement and intersection.  Our principal application is the
transformational approach to negation in higher-order logic programs.
\end{abstract}


\begin{document}
\begin{bottomstuff}
Author's addresses: \newline
A.~Momigliano, Department of Mathematics and Computer Science,
University of Leicester, Leicester, LE1 HR2, U.K.,
\texttt{am133@mcs.le.ac.uk}\newline
F.~Pfenning, Department of Computer Science,
Carnegie Mellon University, Pittsburgh, PA 15213, U.S.A.,
\texttt{fp@cs.cmu.edu} \newline
This work has been support by the National Science Foundation
under grant CCR-9988281.
\end{bottomstuff}

\maketitle

\section{Introduction}
\label{sec:intro}

In most functional and logic programming languages the notion of a
pattern, together with the requisite algorithms for matching or
unification, play an important role in the operational semantics.
Besides unification, other problems such as generalization or complement
also arise frequently.  In this paper we are concerned with the problem
of pattern complement in a setting where patterns may contain binding
operators, so-called \emph{higher-order
patterns}~\cite{Miller91jlc,Nipkow91lics}.  Higher-order patterns have
found applications in logic programming~\cite{Miller91jlc,Pfenning91lf},
logical frameworks~\cite{Despeyroux97}, term
rewriting~\cite{Nipkow93tlca}, and functional logic
programming~\cite{Hanus96RTA}.  Higher-order patterns inherit many
pleasant properties from the first-order case.  In particular, most
general unifiers~\cite{Miller91jlc} and least general
generalizations~\cite{Pfenning91lics} exist, even for complex type
theories.

Unfortunately, the complement operation does not generalize as smoothly.
Lugiez \citeyear{Lug95} has studied the more general problem of
higher-order disunification and had to go outside the language of
patterns and terms to describe complex constraints on sets of solutions.
We can isolate one basic difficulty: a pattern such as $\lambda x\ldot
E\ x$ for an existential variable $E$ matches any term of appropriate
type, while $\lambda x\ldot E$ matches precisely those terms $\lambda
x\ldot M$ where $M$ does not depend on $x$.  The complement then
consists of all terms $\lambda x\ldot M$ such that $M$ \emph{does}
depend on $x$.  However, this set cannot be described by a pattern, or
even a finite set of patterns.

This formulation of the problem suggests that we should consider a
$\lambda$-calculus with an internal notion of \emph{strictness} so
that we can directly express that a term must depend on a given
variable.  For reasons of symmetry and elegance we also add the dual
concept of \emph{invariance} expressing that a given term does
\emph{not} depend on a given variable.  As in the first-order case, it
is useful to single out the case of \emph{linear patterns}, namely
those where no existential variable occurs more than
once.\footnote{This notion of linearity should not be confused with
  the eponymous concept in linear logic and $\lambda$-calculus.}  We
further limit attention to \emph{simple} patterns, that is,  those where
constructors must be strict in their arguments---a condition naturally
satisfied in our intended application domains of functional and logic
programming.  Simple linear patterns in our $\lambda$-calculus of
strict and invariant function spaces then have the following
properties:
\begin{enumerate}
\item The complement of a pattern is a finite set of patterns.
  \item Unification of two patterns is decidable and finitary.
\end{enumerate}
Consequently, finite sets of simple linear patterns in the strict
$\lambda$-calculus are closed under complement and unification.  If we
think of finite sets of linear patterns as representing the set of all
their ground instances, then they form a boolean algebra under
set-theoretic union union, intersection (implemented via unification)
and the complement operation.

The paper is organized as follows: Section~\ref{sec:facomp} briefly
reviews related work and introduces some preliminary definitions.  In
Section~\ref{sec:strict} we introduce a strict $\lam$-calculus and
prove some basic properties culminating in the proof of the existence
of canonical forms in Section~\ref{sec:canonthm}.
Section~\ref{sec:strictcomp} introduces simple terms, followed by the
algorithm for complementation in Section~\ref{sec:compl}.  In
Section~\ref{sec:sunif} we give a corresponding unification algorithm.
Section~\ref{sec:algebra} observes how the set of those patterns can
be arranged in a boolean algebra. We conclude in
Section~\ref{sec:concl} with some applications and speculations on
future research.

\section{Preliminaries and Related Work}
\label{sec:facomp}

A pattern $t$ with free variables can be seen as a representation of the
set of its ground instances, denoted by $\| t \|$.  According to this
interpretation, the \emph{complement} of $t$ is the set of ground terms
that are \emph{not} instances of $t$, i.e., the terms are in the set-theoretic
complement of $\| t \|$.  It is natural to generalize this to
finite sets of terms, where $\|t_1,\ldots,t_n\| = \|t_1\| \cup
\cdots \cup \|t_n\|$.  If we take this one step further we obtain the
important problem of \emph{relative complement}; this corresponds to
computing a suitable representation of all the ground instances of a
given (finite) set of terms which are not instances of another given
one, written as \( \| t_1, \ldots, t_n \| - \| u_1, \ldots, u_m \|. \)

Complement problems have a number of applications in theoretical
computer science (see \citeN{Comon91} for a list of references).  For
example, they are used in functional programming to produce unambiguous
function definitions by patterns and to improve their compilation.  In
rewriting systems they are used to check whether an algebraic
specification is sufficiently complete.  They can also be employed to
analyze communicating processes expressed by infinite transition
systems.  Other applications lie in the areas of machine learning and
inductive theorem proving.  In logic programming, Kunen
\citeyear{Kunen87} used term complement to represent infinite sets of
answers to negative queries.  Our main motivation has been the explicit
synthesis of the negation of higher-order logic programs~[Momigliano
\citeyearNP{Momigliano00phd}; \citeyearNP{Momigliano00csl}], as discussed
briefly in Section~\ref{sec:concl}.

\citeN{Lassez87} proposed the seminal \emph{uncover}
algorithm for computing \emph{first-order} relative complements and
introduced the now familiar restriction to linear terms.  We quote the
definition of the ``$\mnot$'' algorithm for the (singleton)
complement problem given in \cite{Bar90} which we generalize in
Definition~\ref{def:compl}.  Given a finite signature $\Sigma$ and a
linear term $t$ they define:
\[ \begin{array}{lcl}
 \mnot_\Sigma(x) & = & \emptyset \\
 \mnot_\Sigma(f(t_1,\ldots,t_n)) & = &
\{g(x_1,\ldots,x_m) \mid \mbox{$g \in \Sigma$ and $g \not= f$} \} \\
 & \cup & \{f(z_1, \ldots, z_{i-1},s,z_{i+1}, \ldots,z_n) \mid
s\in\mnot_\Sigma(t_{i}), 1\leq i \leq n \}
\end{array} \]
The relative complement problem is then solved by composing the above
complement operation with term intersection implemented via first-order
unification.

An alternative solution to the relative complement problem is
\emph{disunification} (see \cite{Comon91} for a survey and \cite{Lug95}
for an extension to the simply-typed $\lam$-calculus).  Here, operations on
sets of terms are translated into conjunctions or disjunctions of
equations and dis-equations under explicit quantification.
Non-deterministic application of a few dozen rules eventually turns a
given problem into a solved form.  Though a reduction to a significant
subset of the disunification rules is likely to be attainable for
complement problems, control is a major problem.  We argue that using
disunification for this purpose is unnecessarily general.  Moreover, the
higher-order case results in additional complications, such as restrictions on
the occurrences of bound variables, which fall outside an otherwise
clean framework.  As we show in this paper, this must not necessarily be
the case.  We believe that our techniques can also be applied to analyze
disunification, although we have not investigated this possibility at
present.

We now introduce some preliminary definitions and examples which guide
our development.  We begin with the simply-typed $\lambda$-calculus.  We
write $a$ for atomic types, $c$ for term-level constants, and $x$ for
term-level variables.  Note that variables $x$ should be seen as
parameters and not subject to instantiation.

\[ \begin{array}{rrcl}
 \mbox{\textit{Simple Types}} & A & \bnfas & a \bnfalt A_1 \arrow A_2 \\
 \mbox{\textit{Terms}} & M & \bnfas & c \bnfalt x \bnfalt \lam
 x\oftp A\ldot M \bnfalt M_1\ M_2 \\ 
 \mbox{\textit{Signatures}} & \Sigma & \bnfas & \cdot \bnfalt
 \Sigma, a \oftp \mbox{type} \bnfalt \Sigma, c\oftp A \\
 \mbox{\textit{Contexts}} & \GG & \bnfas & \cdot \bnfalt \GG, x\oftp A
\end{array} \]

We require that signatures and contexts declare each constant or variable at
most once.  Furthermore, we identify contexts that differ only in their
order and promote `,' to denote disjoint set union.  As usual we identify
terms which differ only in the names of their bound variables.  We restrict
attention to well-typed terms, omitting the standard typing rules.
We write the main typing judgment as $\Gamma \vd M \hastype A$, assuming
a fixed signature $\Sigma$.

In applications such as logic programming or logical frameworks,
$\lambda$-abstraction is used to represent binding operators in some
object language.  In such a situation the most appropriate notion of
normal form is the long $\beta\eta$-normal form (which we call
\emph{canonical form}), since canonical forms are almost always the
terms in bijective correspondence with the objects we are trying to
represent.  Every well-typed term in the simply-typed $\lambda$-calculus
has a unique canonical form---a property which persists in the strict
$\lambda$-calculus introduced in Section~\ref{sec:strict}.

We denote existential variables of type $A$ (also called logical variables,
meta-variables, or pattern variables) by $E_A$, although we mostly omit the
type $A$ when it is clear from the context.  We think of existential variables
as syntactically distinct from bound variables or free variables declared in a
context.  A term possibly containing some existential variables is called a
\emph{pattern} if each occurrence of an existential variable appears in a
subterm of the form $E\ x_1\ldots x_n$, where the arguments $x_i$ are distinct
occurrences of free or bound variables (but not existential variables).  We
call a term \emph{ground} if it contains no existential variables.  Note that
it may still contain parameters.

Semantically, an existential variable $E_A$ stands for all canonical
terms $M$ of type $A$ in the empty context with respect to a given
signature.  We extend this to arbitrary well-typed patterns in the usual
way, and write $\GG \vd M\in\| N\| \hastype A$ when a term $M$ is a instance
of a pattern $N$ at type $A$ containing only the parameters in $\GG$
and no existential variables.  In this setting, unification of two
patterns without shared existential variables corresponds to an
intersection of the set of terms they
denote~\cite{Miller91jlc,Pfenning91lics}.  This set is always either
empty, or can be expressed again as the set of instances of a single
pattern.  That is, patterns admit most general unifiers.

The class of higher-order patterns inherits many properties from
first-order terms.  However, as we will see, it is \emph{not} closed
under complement, but a special subclass is.  We call a canonical
pattern $\GG \vd M \hastype A$ \emph{fully applied} if each
occurrence of an existential variable $E$ under binders $y_1,\ldots,y_m$
is applied to some permutation of the variables in $\GG$ and
$y_1,\ldots,y_m$.  Fully applied patterns play an important role in
functional logic programming and rewriting~\cite{Hanus96RTA}, because any
fully applied existential variable $\GG \vd E\ x_1\ldots x_n \hastype
a$ denotes all canonical terms of type $a$ with parameters from
$\GG$.  It is this property which makes complementation particularly
simple.

\begin{example}
\label{ex:lam}
Consider the untyped $\lam$-calculus:\footnote{We
use $\Lambda$ and $@$ to avoid confusion with $\lam$ and application in
the language of patterns.}
 \begin{eqnarray*}
  e & \bnfas & x \mid \Lambda x\ldot e \mid e_1\ @\  e_2
 \end{eqnarray*}
We encode these expressions using the usual technique of higher-order abstract
syntax as canonical forms over the following signature.
 \begin{eqnarray*}
  \Sigma  & = & \itexp\hastype \mbox{\textrm{type}}, \itlam\hastype (\itexp\arrow
  \itexp)\arrow \itexp, \itapp \hastype \itexp \arrow \itexp\arrow \itexp
 \end{eqnarray*}
 The representation function $\rep{\_}$ is defined as follows:
 \begin{eqnarray*}
  \rep{x} & = & x\hastype \itexp\\
  \rep{\Lambda x\ldot e} & = & \itlam\ (\lamb{x}{\itexp} \rep{e})\\
  \rep{e_1\ @ \ e_2} & = & \itapp\ \rep{e_1}\ \rep{e_2}
  \end{eqnarray*}
 The representation of an object-language $\beta$-redex then has the
form
 \begin{eqnarray*}
  \rep{(\Lambda x\ldot e)\ @\ f} & = & \itapp\ (\itlam\  (\lamb{x}{\itexp} \rep{e}))\ \rep{f},
 \end{eqnarray*}
 where $\rep{e}$ may have free occurrences of $x$.  When written as a
pattern with existential variables $E_{\itexp \arrow \itexp}$ and $F_{\itexp}$
this is expressed as
 \[ \itapp\ (\itlam\ (\lamb{x}{\itexp} E\ x)\ F). \]
 Note that in the empty context this pattern is fully applied.  Its
complement with respect to the empty context contains every top-level
abstraction plus every application where the first argument is not an
abstraction.
 \[ \mnot(\itapp\ (\itlam\ (\lamb{x}{\itexp} E\ x)\ F)) = 
    \{\itlam\ (\lamb{x}{\itexp} H\ x), \itapp\ (\itapp\ H_{1}\ H_{2})\ H_{3}\}
 \]
 Here $H$, $H_1$, $H_2$, $H_3$ are fresh existential variables of appropriate
type, namely $H \hastype \itexp \arrow \itexp$ and $H_i \hastype \itexp$.
\end{example}

For patterns that are not fully applied, the complement cannot
be expressed as a finite set of patterns, as the following example
illustrates.

\begin{example}
 \label{ex:etardx}
 The encoding of an $\eta$-redex takes the form:
 \begin{eqnarray*}
  \rep{\Lambda x.\, (e\ @\ x)} & = & \itlam\ (\lamb{x}{\itexp} \itapp\ \rep{e}\ x)
 \end{eqnarray*}
 where $\rep{e}$ may contain no free occurrence of $x$.  The side
condition is expressed in a pattern by introducing an existential
variable $E_{\itexp}$ which {\em does not\/} depend on $x$, that is
 \[ \itlam\ (\lamb{x}{\itexp} \itapp\ E\ x). \]
  Hence, its complement with respect to the empty context should
contain, among others, also all terms
 \[ \itlam\ (\lamb{x}{\itexp} \itapp\ (F\ x)\ (H\ x)) \] where
$F\hastype \itexp \arrow \itexp$ {\em must\/} depend on its argument $x$
while $H\hastype \itexp \arrow \itexp$ may or may not depend on $x$.
\end{example}

As the example above shows, the complement of patterns that are not fully
applied cannot be represented as a finite set of patterns.  Indeed, there is
no finite set of patterns which has as its ground instances exactly those
terms $M$ which depend on a given variable $x$.  This failure of closure under
complementation cannot be avoided similarly to the way in which
left-linearization bypasses the limitation to linear patterns and it needs to
be addressed directly.

One approach is taken by \citeN{Lug95}: he modifies the language of
terms to permit occurrence constraints.  For example $\lam xyz.\ M\{1,3\}$
would denote a function which depends on its first and third argument.  The
technical handling of those objects then becomes awkward as they require
specialized rules which are foreign to the issues of complementation.

Since our underlying $\lambda$-calculus is typed, we use typing to
express that a function \emph{must} depend on a variable $x$. Following
standard terminology, we call such terms \emph{strict in $x$} and the
corresponding function $\lamb{x}{A} M$ a \emph{strict function}. In the
next section we develop such a $\lambda$-calculus and then generalize the
complement algorithm to work on such terms.

\section{Strict Types}
\label{sec:strict}

As we have seen in the preceding section, the complement of a partially
applied pattern in the simply-typed $\lambda$-calculus cannot be
expressed in a finitary manner within the same calculus.  We thus
generalize our language to include \emph{strict} functions of type $A
\RImp B$ (which are guaranteed to depend on their argument) and
\emph{invariant} functions of type $A \ZImp B$ (which are guaranteed
\emph{not} to depend on their argument).  Of course, any concretely
given function either will or will not depend on its argument, but in
the presence of higher-order functions and existential variables we
still need the ability to remain uncommitted.  Therefore our calculus
also contains the full function space $A \TImp B$.  We first concentrate
on a version without existential variables.  A similar calculus has been
independently investigated by \citeN{Wright92} and \citeN{Baker94}; for
a comparison see the end of Section~\ref{sec:canonthm}.

\[ \begin{array}{rrcl}
 \mbox{\textit{Labels}} & k & \bnfas & 1 \bnfalt 0 \bnfalt u \\
 \mbox{\textit{Types}} & A & \bnfas & a \bnfalt A_1 \GImp{k} A_2 \\
 \mbox{\textit{Terms}} & M & \bnfas & c \bnfalt x \bnfalt \lam
 x^k\oftp A\ldot M \bnfalt M_1\ M_2^k
 \end{array} \]

Note that there are three different forms of abstractions and
applications, where the latter are distinguished by different labels on
the argument.  It is not really necessary to distinguish three forms of
application syntactically, since the type of a function determines the
status of its argument, but it is convenient for our purposes.  A label
$u$ is called \emph{undetermined}, otherwise it is \emph{determined} and
denoted by $d$.

We use a formulation of the typing judgment
  \[ \GG; \OO; \DD \vd M \hastype A \]
 with three zones: $\GG$ containing \emph{unrestricted} hypotheses,
$\OO$ containing the \emph{irrelevant} hypotheses, and $\DD$ containing
the \emph{strict} hypotheses.  We implicitly assume a fixed signature
$\Sigma$ which would otherwise clutter the presentation.  Recall that
$\GG_1,\GG_2$ is the union of two contexts that do not declare any
common variables.  Recall also that we consider contexts as sets, that
is, exchange is left implicit.  The typing rules are given in
Figure~\ref{fig:typrules}.

\begin{figure}

\[ \begin{array}{c}
 \ianc{c\oftp A \in \Sigma}{\GG; \OO; \cdot \vd c \hastype
 A}{\mathtt{Con}} \\[1em]

 \ianc{}{(\GG, x\oftp A); \OO; \cdot \vd x \hastype A}{\mathtt{Id}^u}
 \hspace{2em} \mbox{\textit{no $\mathtt{Id}^0$ rule}} \hspace{2em}
 \ianc{}{\GG; \OO; x\oftp A \vd x \hastype A}{\mathtt{Id}^1} \\[1em]

 \ianc{(\GG, x\oftp A); \OO; \DD \vd M \hastype B}{\GG;
\OO; \DD \vd \lam x^u \oftp A\ldot M \hastype A \GImp{u}
B}{\GImp{u}I} \\[1em]

 \ianc{\GG; (\OO, x\oftp A); \DD \vd M \hastype B}{\GG;
\OO; \DD \vd \lam x^0 \oftp A\ldot M \hastype A \GImp{0}
B}{\GImp{0}I} \\[1em]

 \ianc{\GG; \OO; (\DD, x\oftp A)\vd M \hastype B}{\GG;
\OO; \DD \vd \lam x^1 \oftp A\ldot M \hastype A \GImp{1}
B}{\GImp{1}I} \\[1em]

 \ibnc{\GG; \OO; \DD \vd M \hastype A \TImp B}{(\GG,
\DD); \OO; \cdot \vd N \hastype A}{\GG; \OO; \DD \vd M\
N^u \hastype B}{\GImp{u}E} \\[1em]

 \ibnc{\GG; \OO; \DD \vd M \hastype A \ZImp B}{(\GG,
\OO,\DD); \cdot; \cdot \vd N \hastype A}{\GG; \OO; \DD
\vd M\ N^0 \hastype B}{\GImp{0}E} \\[1em]

\ibnc{(\GG, \DD_N); \OO; \DD_M \vd M \hastype A \RImp
B}{(\GG, \DD_M); \OO; \DD_N \vd N \hastype A}{\GG;
\OO; (\DD_M, \DD_N) \vd M\ N^1 \hastype B}{\GImp{1}E}

 \end{array} \]
\caption{Typing rules for $\god\vd M\hastype A$}
\label{fig:typrules}
\end{figure}

Our system is biased towards a bottom-up reading of the rules in that
variables never disappear, i.e., they are always propagated from the
conclusion to the premises, although their status might be changed.

Let us go through the typing rules in detail.  The requirement for the
strict context $\DD$ to be empty in the $\mathtt{Id}^u$ and
$\mathtt{Id}^1$ rules expresses that strict variables must be used,
while undetermined variables in $\GG$ or irrelevant variables in
$\OO$ can be ignored.  Note that there is no rule for irrelevant
variables, which expresses that they cannot be used.  The introduction
rules for undetermined, invariant, and strict functions simply add a
variable to the appropriate context and check the body of the function.
The difficult rules are the three elimination rules.  First, the
unrestricted context $\GG$ is always propagated to both premises.
This reflects that we place no restriction on the use of these
variables.

Next we consider the strict context $\DD$: recall that this contains
the variables which should occur strictly in a term.  An undetermined
function $M\hastype A \GImp{u} B$ may or may not use its argument.  An
occurrence of a variable in the argument to such a function can
therefore not be guaranteed to be used.  Hence we must require in the
rule $\GImp{u}E$ for an application $M\ N^u$ that all variables in
$\DD$ occur strictly in $M$.  This ensures at least one strict
occurrence in $M$ and no further restrictions on occurrences of strict
variables in the argument are necessary.  This is reflected in the rule
by adding $\DD$ to the unrestricted context while checking the
argument $N$.  The treatment of the strict variables in the vacuous
application $M\ N^0$ is similar.

In the case of a strict application $M\ N^1$ each strict variable
should occur strictly in either $M$ or $N$.  We therefore split the
context into $\DD_M$ and $\DD_N$ guaranteeing that each variable
has at least one strict occurrence in $M$ or $N$, respectively.
However, strict variables can occur more than once, so variables from
$\DD_N$ can be used freely in $M$, and variables from $\DD_M$
can occur freely in $N$.  As before, we reflect this by adding these
variables to the unrestricted context.

Finally we consider the irrelevant context $\OO$.  Variables
declared in $\OO$ cannot be used \emph{except} in the argument to
an invariant function (which is guaranteed to ignore its argument).
We therefore add the irrelevant context $\OO$ to the unrestricted
context when checking the argument of a vacuous application $M\ N^0$.

We now illustrate how the strict application rule
non-deterministically splits contexts. Consider the typing problem
$\cdot;\cdot; (x \oftp A \RImp A\RImp B ,y\oftp A)\vd (x\ y^1) \ 
y^1\hastype B$, related to the contraction principle.  There are four
ways to split the strict context for the outer application.

\[\begin{array}{ll}
\DD_{M} = x \oftp A \RImp A\RImp B ,y\oftp A & \DD_{N}=\cdot\\ \DD_{M} =
x \oftp A \RImp A\RImp B & \DD_{N} = y\oftp A\\ \DD_{M} = y\oftp A &
\DD_{N}= x \oftp A \RImp A\RImp B\\ \DD_{M} =\cdot & \DD_{N}=x \oftp A
\RImp A\RImp B ,y\oftp A
\end{array}\]
 Only the first two yield a valid derivation as depicted in
Figures~\ref{fig:strictex} and~\ref{fig:strictex2}.  Here we
have dropped the types in the context.

\begin{figure}
 \newcommand{\aaa}{y; \cdot; x  \vd x \hastype A\RImp A\RImp B}
 \newcommand{\bbb}{x;\cdot ; y \vd y\hastype A }
 \newcommand{\ccc}{\cdot;\cdot ;( x  , y  )\vd x\ y^1 \hastype A\RImp B}
 \newcommand{\ddd}{(x, y ); \cdot ;\cdot\vd y\hastype A}
 \newcommand{\eee}{\cdot;\cdot; (x \oftp A \RImp A\RImp B, y\oftp
  A)\vd (x\ y^1) \ y^1 \hastype B}
\[ \ianc{\ian{\ian{}{\aaa}{\mathtt{Id}^1} \hspace{2em}
           \ian{}{\bbb}{\mathtt{Id}^1}}
          {\ccc}{{\RImp}E} \hspace{7em}
      \ian{}{\ddd}{\mathtt{Id}^u}}
     {\eee}{{\RImp}E}
\]
\caption{First derivation of $\cdot;\cdot; (x  \oftp A \RImp A\RImp B
  ,y\oftp A) \vd (x\ y^1)\  y^1\hastype B$} 
\label{fig:strictex}
\end{figure}

\begin{figure}
  \newcommand{\aaa}{y ; \cdot; x \vd x \hastype A \RImp A\RImp B}
  \newcommand{\bbb}{(x , y );\cdot;\cdot\vd y\hastype A }
  \newcommand{\ccc}{y ;\cdot ; x \vd x\ y^1 \hastype A\RImp B}
  \newcommand{\ddd}{ x ;\cdot; y \vd y\hastype A}
  \newcommand{\eee}{
    \cdot;\cdot; (x \oftp A \RImp A\RImp B ,y\oftp A)\vd (x\ y^1) \ y^1
    \hastype B}
 \[ \ianc{\ian{\ian{}{\aaa}{\mathtt{Id}^1} \hspace{2em}
           \ian{}{\bbb}{\mathtt{Id}^u}}
          {\ccc}{{\RImp}E} \hspace{9em}
      \ian{}{\ddd}{\mathtt{Id}^1}}
     {\eee}{{\RImp}E}
 \]
\caption{Second derivation of  $\cdot;\cdot; (x  \oftp A \RImp A\RImp
  B ,y\oftp A)\vd (x \ y^1)\  y^1\hastype B$} 
\label{fig:strictex2}
\end{figure}

Our strict $\lambda$-calculus satisfies the expected properties,
culminating in the existence of canonical forms which is critical for
the intended applications.  First we remark that types are unique,
although typing derivations may not.

\begin{theorem}[Uniqueness of Typing] \mbox{}
\label{thm:uniqueT}
  Assume $(\GG,\OO,\DD) = (\GG',\OO',\DD')$. \newline
  If $\GG;\OO;\DD \vd M\hastype A$ and $\GG';\OO';\DD'\vd M\hastype A'$,
 then $A = A'$.
\end{theorem}
\begin{proof}
  By induction on the structure of the given derivation, exploiting
  uniqueness for declarations of variables and constants.
\end{proof}

We start addressing the structural properties of the contexts.  Exchange
is directly built into the formulation and will not be repeated.  Note
that our calculus is formulated entirely without structural rules, which
now have to be shown to be admissible.

\begin{lemma}[Weakening] \mbox{}
\label{thm:weak}
\begin{enumerate}
\item (Weakening$^{u}$) If $\GG; \OO; \DD\vd M \hastype A$,
  then $(\GG, x\oftp C); \OO; \DD\vd M \hastype A$.
\item (Weakening$^{0}$) If $\GG; \OO; \DD\vd M \hastype A$,
  then $\GG ; (\OO, x\oftp C); \DD\vd M \hastype A$.
\end{enumerate}
\end{lemma}
 
\begin{proof}
  By induction on the structure of the given derivations.
\end{proof}

The following properties allow us to lose track of strict and vacuous
occurrences, if we are so inclined.

\begin{lemma}[Loosening]
\label{thm:loosex}
\mbox{}
\begin{enumerate}
\item (Loosening$^{0}$)
 If $\GG; (\OO, x\oftp C); \DD \vd M \hastype A$, then $(\GG,x\oftp
C);\OO;\DD \vd M \hastype A$.
\item (Loosening$^{1}$)
 If $\GG; \OO; (\DD,x\oftp C)\vd M \hastype A$, then $(\GG, x\oftp C);
\OO; \DD \vd M \hastype A$.
\end{enumerate}
\end{lemma}
\begin{proof}
By induction on the structure of the given derivations.
\end{proof}

Next we come to the critical substitution properties.  They verify the
intended meaning of the hypothetical judgments and directly entail
subject reduction (Theorem~\ref{thm:subred}).  To be consistent with the
design of our typing rules, we formulate the substitution properties so
that each of the given derivation depends on the same variables,
although their status might be different (unrestricted, irrelevant, or
strict).  Note that this is possible only because we have included
irrelevant hypotheses in our judgment.

\begin{lemma}[Substitution] \mbox{}
\label{thm:substitution}
\begin{enumerate}
\item \label{part:subu}
  (Substitution$^u$) If $(\GG, x\oftp A);\OO;\DD\vd M
  \hastype C$ and $(\GG,\DD); \OO;\cdot \vd N \hastype A$, then
  $\GG;\OO;\DD \vd [N/x]M \hastype C$.
\item (Substitution$^0$) If $\GG;(\OO,x\oftp A);\DD \vd M
  \hastype C$ and $(\GG, \DD,\OO) ;\cdot;\cdot \vd N \hastype
  A$, then $\GG;\OO;\DD \vd [N/x]M \hastype C$.  
\item \label{part:sub1}
  (Substitution$^1$) If $(\GG,\DD_N);\OO; (\DD_M,
  x\oftp A )\vd M \hastype C$ and $(\GG,\DD_M);\OO;\DD_N
  \vd N\hastype A$, then
  $\GG; \OO;( \DD_M, \DD_N) \vd [N/x]M \hastype C$.
\end{enumerate}
\end{lemma}

\begin{proof}
 We proceed by mutual induction on the structure of the derivation $\cD$ of
$M\hastype C$, using weakening and loosening as needed to match the form of
the induction hypothesis.  Each case is otherwise entirely straightforward.
We show only one case in the proof of strict substitution
(part~\ref{part:sub1}).  Here and in subsequent proofs we sometimes write $\cD
:: J$ if $\cD$ is a derivation of judgment $J$ instead of the two-dimensional
notation $\arv{\cD}{J}$.

\begin{description}
 \item[Case] $\cD$ ends in ${\RImp}E$.  There are two sub-cases, depending
on whether the declaration $x\oftp A$ is strict in the left premise or
right premise.  We show the former.
 \[ \ianc{\arv {\cD_1}{(\GG,\DD_{N},\DD_{Q}); \OO; 
(\DD_{P}, x\oftp A) \vd P \hastype B \RImp C} \hspace{1em} \arv{\cD_2}{ 
(\GG,\DD_{N},\DD_{P}, x\oftp A);\OO;\DD_{Q} \vd Q \hastype B} }
{(\GG,\DD_N); \OO; (\DD_{P}, x\oftp A,\DD_{Q})\vd P\ Q^1 \hastype
  C}{\GImp{1}E} \]

\begin{tabbing}
 $\cD_1::(\GG,\DD_{N},\DD_{Q}); \OO; (\DD_{P}, x\oftp A) \vd P
\hastype B \RImp C$ \` Subderivation\\
 $\cE::(\GG,\DD_P, \DD_{Q});\OO;\DD_{N}\vd N\hastype A$ \` Assumption \\
 $(\GG,\DD_{Q}); \OO; (\DD_{P}, \DD_N) \vd [N/x] P \hastype B \RImp
C$ \` By i.h. (\ref{part:sub1}) on $\cD_{1}, \cE$\\
 $(\GG,\DD_{Q},\DD_N); \OO; \DD_{P} \vd [N/x] P \hastype B
\RImp C$ \` By Loosening$^1$ $\DD_N$\\
 $\cD_2::(\GG,\DD_{N}, \DD_{P}, x\oftp A);\OO;\DD_{Q} \vd Q \hastype B$ \` 
Subderivation\\
 $\cE'::(\GG,\DD_P,\DD_Q,\DD_N);\OO;\cdot\vd N\hastype A$ \`By
 Loosening$^{1}$ $\DD_N$ in $\cE$\\
 $(\GG,\DD_{N},\DD_{P}); \OO; \DD_{Q} \vd [N/x] Q \hastype B$ \` By 
i.h. (\ref{part:subu}) on $\cD_{2}, \cE'$\\
$\GG;\OO;(\DD_{P},\DD_{Q},\DD_{N})\vd[N/x] (P\ Q^{1})\hastype C$\` By rule
 ${\RImp}E$
\end{tabbing}
\end{description}
\end{proof}

Weakening, loosening, and substitution directly imply the contraction
property for all three kinds of hypotheses.  Since we do not use
contraction in this paper, we elide the formal statement and proof
of this property.

The notions of reduction and expansion derive directly from the ordinary
$\beta$ and $\eta$ rules.

\[ \begin{array}{rcl}
   (\lam x^k\oftp A\ldot M)\, N^k & \betared & [N/x]M \\[1ex]
   (M \hastype A \GImp{k} B) & \etaexp & \lam x^k\oftp A\ldot M\  x^k
\end{array} \]

An application of $\eta$-expansion rules requires the term $M$ to have
the indicated type.  The subject reduction and expansion theorems are an
immediate consequence of the structural and substitution properties.

\begin{theorem}[Subject Reduction] \mbox{} \newline
\label{thm:subred}
 If $\GG;\OO; \DD \vd M \hastype A$ and $M \betared M'$ then 
$\GG;\OO; \DD \vd M' \hastype A$.
\end{theorem}
\begin{proof} We proceed by cases and inversion followed
by an appeal to the substitution property.  We show only one case.
Let $M = (\lamb{x^{1}}{B} P)\ Q^{1}\hastype A$ and $M' = [Q/x]P$.  
 \begin{tabbing}
  $\GG;\OO;\DD\vd (\lamb{x^{1}}{B} P)\ Q^{1}\hastype A$ \` Assumption \\
  $\DD=(\DD_{P},\DD_{Q})$, $\cE::(\GG,\DD_{P});\OO;\DD_{Q}\vd Q\hastype B$, and \\
 $(\GG,\DD_{Q});\OO;\DD_{P}\vd \lamb{x^{1}}{B} P\hastype B\RImp A$ \` By inversion \\
 $\cD::(\GG,\DD_{Q});\OO;(\DD_{P},x\oftp B)\vd P\hastype A$
 \` By further inversion \\
 $\GG;\OO;(\DD_{P},\DD_{Q})\vd[Q/x]P\hastype A$
 \`By substitution$^{1}$ on $\cD,\cE$ 
\end{tabbing}
\end{proof}

 Subject reduction continues to hold if we allow the reduction of an arbitrary
subterm occurrence.  We omit the obvious statement and formal proof of
this fact.

\begin{theorem}[Subject Expansion] \mbox{} \newline
\label{thm:subexp}
 If $\GG;\OO; \DD \vd (M \hastype A) \GImp{k} B$ and $(M \hastype
A\GImp{k} B) \etaexp M'$ then $\GG;\OO; \DD \vd M' \hastype A
\GImp{k} B$.
\end{theorem}
\begin{proof} Direct.  We consider only the strict case ($k = 1$).
 \begin{tabbing}
  $\GG;\OO;\DD\vd M\hastype A\RImp B$ \` Assumption \\
  $(\GG,x\oftp A);\OO; \DD\vd M \hastype A\RImp B$\` By
 weakening$^{u}$ \\
  $(\GG,\DD);\OO;x\oftp A\vd x\hastype A$ \` By rule $\mathtt{Id}^{1}$ \\
  $\GG;\OO; (\DD,x\oftp A) \vd M\ x^{1} \hastype B$
  \` By rule ${\RImp}E$ \\
  $\GG;\OO; \DD\vd \lamb{x^{1}}{A} M\ x^{1} \hastype A\RImp B$
  \` By rule $\RImp I$
 \end{tabbing}
\end{proof}

The following lemma establishes a sort of consistency property of the
type system, showing that a term $M$ cannot be both strict and vacuous
in a given variable.  This will be central in the proof of disjointness
of pattern complementation (Lemma~\ref{le:part1}).

\begin{lemma}[Exclusivity] \mbox{}
\label{le:disj}
It is not the case that both $\GG_1;\OO_1;(\DD_1,x\oftp C)\vd M\hastype 
A$ and $\GG_2;(\OO_2,x\oftp C);\DD_2\vd M\hastype A$.
\end{lemma}
\begin{proof}
  By induction on the structure of the derivation of
$\GG_1;\OO_1;(\DD_1,x\oftp C)\vd M\hastype A$, applying inversion on the
derivation of $\GG_2;(\OO_2,x\oftp C);\DD_2\vd M\hastype A$ in each
case.
\end{proof}

\section{The Canonical Form Theorem}
\label{sec:canonthm}

In this section we establish the existence of canonical forms for the
strict $\lambda$-calculus, i.e., $\beta$-normal $\eta$-long forms, which
is crucial for our intended application. We prove this by Tait's
method of \emph{logical relations}; we essentially follow the account
in \cite{Pfenning01book}, with a surprisingly little amount of
generalization from simple to strict types, thanks to  a
simplified account of substitutions.

We start by presenting the inductive definition of \emph{canonical}
forms.  It is realized by the two mutually recursive
judgments depicted in Figure~\ref{fig:canon}:
 \[ \begin{tabular}{ll}
   $\GG;\OO;\DD\vd M \da A$ & $M$ is atomic of type $A$. \\
   $\GG;\OO;\DD\vd M \bua A$ & $M$ is canonical of type $A$.
 \end{tabular} \]

\begin{figure}
\[
\begin{array}{c}
\ianc{c\oftp A\in\Sigma}{\GG; \OO;\cdot  \vd  c\da A}{\mathtt{cIdc}}\\[2ex]
\ianc{}{(\GG,x\oftp A);\OO; \cdot  \vd  x\da A}{\mathtt{cId}^{u}}
\qquad \mbox{\textrm{no $\mathtt{cId}^0$ rule}}
\qquad
\ianc{}{\GG;\OO; x\oftp A \vd  x\da A} {\mathtt{cId}^{1}}
\\[2ex]
%\ianc{}{\GG;\OO;\DD\vd E\ \ol{x_{n}^{l}}\da P}{cPat}
%\qquad
\ianc{\GG;\OO;\DD \vd  M\da a}{\GG;\OO;\DD \vd  M\bua a} 
{\mathtt{cAt}}\\[2ex]
\ianc{(\GG,x\oftp A) ; \OO;\DD \vd M\bua B}{\GG;\OO;\DD\vd(\lamb{x^u}{A} M) \bua 
A\TImp B} {\mathtt{c}\TImp I}
\\[2ex]
\ianc{\GG ; \OO; (\DD,x\oftp A) \vd M\bua B}{\GG;\OO;\DD\vd(\lamb{x^1}{A} M) \bua 
A\RImp B}{\mathtt{c}\RImp I}
\\[2ex]
\ianc{\GG ;(\OO, x\oftp A) ;\DD \vd M\bua B}{\GG;\OO;\DD\vd(\lamb{x^0}{A} M) \bua 
A\ZImp B}{\mathtt{c}\ZImp I}
\\[2ex]
 \ibnc{\GG; \OO; \DD \vd M \da A \TImp B}{(\GG, \DD);
\OO; \cdot \vd N \bua A}{\GG; \OO; \DD
\vd M\ N^u \da B}{\mathtt{c}\GImp{u}E}
\\[2ex]
\ianc{\GG;\OO;\DD\vd M \da A \ZImp B \quad (\GG,\OO,\DD);\cdot;\cdot \vd N \bua A
} {\GG;\OO;\DD \vd M \ N^0 \da B}{\mathtt{c}{\ZImp}E}\\[2ex]
\ibnc{(\GG, \DD_N); \OO; \DD_M \vd M \da A \RImp 
B}{(\GG, \DD_M); \OO; \DD_N \vd N \bua A}{\GG; \OO;
(\DD_M, \DD_N) \vd M\ N^1 \da B}{\mathtt{c}{\RImp}E}
\end{array}
\]
\caption{Canonical forms: $\god\vd M\buda A$}
\label{fig:canon}
\end{figure}

\begin{figure}
\[
\begin{array}{c}
\ianc{c\oftp A\in\Sigma}{\Psi \vd  c\da c\hastype A}{\mathtt{tcIdc}}\qquad
%\ianc{}{\Psi\vd  x\da x\hastype A}{tcId^{u}}
%\qquad \mbox{\textrm{no rule for }}tcId^{0}
\qquad
\ianc{x\oftp A\in\Psi}{\Psi \vd  x\da x\hastype A} {\mathtt{tcIdvar}}
\\[2ex]
\ibnc{M\whr M'}{\Psi\vd M'\bua M''\hastype a}{\Psi\vd M\bua M''\hastype a}{\mathtt{tc}
\whr}
\qquad
\ianc{\Psi \vd  M\da N\hastype a}{\Psi\vd
  M\bua N\hastype a}{\mathtt{tcAtm}}
\\[2ex]
\ianc{\Psi,x\oftp A \vd M\ x^k\bua N\hastype B}
{\Psi\vd M\bua (\lamb{x^x}{A} N) \hastype A\GImp{k} B}
{\mathtt{tc}\GImp{k} I}
\qquad
% \ianc{\GG ;(\OO, x\oftp A) ;\DD \vd M\ x^0\bua N\hastype B}
% {\GG;\OO;\DD\vd M\bua (\lamb{x^0}{A} N) \hastype A\ZImp B}{\mathtt{tc}\ZImp I}
% \\[2ex]
% \ianc{\GG ; \OO; (\DD,x\oftp A) \vd M\ x^1\bua N\hastype B}
% {\GG;\DD\vd M\bua(\lamb{x^1}{A} N) \hastype A\RImp B}{\mathtt{tc}\RImp I}
% \\[2ex]
 \ibnc{\Psi\vd M \da P\hastype A \GImp{k} B}{\Psi \vd N \bua Q
 \hastype A}{\Psi \vd M\ N^k \da P\ Q^k\hastype B}{\mathtt{tc}\GImp{k}E}
% \\[2ex]
% \ibnc{\GG;\OO;\DD\vd M \da P\hastype A \ZImp B}{ (\GG,\OO,\DD);
% \cdot;\cdot \vd N \bua Q\hastype A}
% {\GG;\OO;\DD \vd M \ N^0 \da P\ Q^0\hastype B}{\mathtt{tc}{\ZImp}E}\\[2ex]
% \ibnc{(\GG, \DD_N); \OO; \DD_M \vd M \da P\hastype A \RImp 
% B}{(\GG, \DD_M); \OO; \DD_N \vd N \bua Q\hastype
% A}{\GG; \OO; (\DD_M, \DD_N) \vd M\ N^1 \da P\ Q^1\hastype
% B}{\mathtt{tc}{\RImp}E}
\end{array}
\]
\caption{Conversion to canonical form: $\Psi\vd M \buda N\hastype A$}
\label{fig:tocanon}
\end{figure}

Informally, $M$ is atomic (written $M \da A$ for some $A$) if $M$
consists of a variable applied to a sequence of arguments, where each of
the arguments is canonical at appropriate type.  A term $M$ is canonical
if $M$ consists of a sequence of $\lambda$-abstractions followed by an
atomic term of atomic type.  We shall abbreviate judgments involving
$\bua$ and $\da$ as $\buda$.

\begin{lemma}[Soundness of  Canonical Terms] \mbox{} \newline
\label{le:conversound}
If $\god\vd M\buda A$, then $\god\vd M\hastype A$.
\end{lemma}
\begin{proof}
 By induction on the structure of the derivation of $\god\vd
  M\buda A$.
\end{proof}

We describe an algorithm for conversion to canonical form in
Figure~\ref{fig:tocanon}.  This algorithm is presented as a deductive
system that can be used to construct a canonical form from an arbitrary
well-typed term.  Note that the algorithm does not need to keep track of
occurrence constraints---they will be satisfied by construction (see
Theorem~\ref{thm:convercf}).  We write $\Psi$ for a single context of
distinct variable declarations whose status should be considered
ambiguous since it is unnecessary to know whether they are unrestricted,
irrelevant, or strict.

\[ \begin{tabular}{ll}
 $\Psi\vd M \da N\hastype A$ & $M$ has atomic form $N$ of type $A$. \\
 $\Psi\vd M \bua N\hastype A$ & $M$ has canonical form $N$ at type $A$.
\end{tabular} \]
 These utilize weak head reduction, which
includes local reduction ($\beta$) and partial congruence ($\nu$):
\begin{eqnarray*}
 \ianc{}{(\lamb{x^k}{A} M)\ N^k \whr [N/x]M}{\beta^k}\qquad
 \ianc{M\whr Q}{M\ N^k \whr Q\ N^k}{\nu^k}
\end{eqnarray*}
  Operationally, we assume that $M$ is given and we construct
an $N$ such that $M \whr N$ or fail.  The judgments for conversion
to canonical form can be interpreted as an algorithm in the following
manner:
\[ \begin{tabular}{ll}
  $\Psi \vd M \da N \hastype A$ & Given $\Psi$ and $M$ construct $N$ and $A$ \\
  $\Psi \vd M \bua N \hastype A$ & Given $\Psi$, $M$, and $A$ construct $N$ \\
 \end{tabular} \]
 The main theorem of this section states that if $\GG; \OO; \DD \vd
M \hastype A$ and $\Psi = (\GG,\OO,\DD)$ then the two judgments
above will always succeed to construct an $N$ and $A$, or $N$, respectively.

\begin{theorem}[Conversion Yields Canonical Terms]\mbox{} \newline
\label{thm:convercf}
 If $(\GG,\OO,\DD) \vd M\buda N\hastype A$ and $\GG; \OO;
\DD \vd M\hastype A$, then $\god\vd N\buda A$.
\end{theorem}
\begin{proof}
 By induction on the structure of the derivation of
$(\GG,\OO,\DD) \vd M\buda N\hastype A$ and inversion on the
given typing derivation in each case.
\end{proof}

In the construction of logical relations we will need a notion of
context extension, $\Psi' \geq \Psi$ ($\Psi'$ extends $\Psi$ with zero
or more declarations).  It is clear that conversion to canonical form is
not affected by weakening.  We omit the formal statement of this
property.

We can now introduce a unary \emph{Kripke-logical relation}, in complete
analogy with the usual definition for the simply-typed $\lam$-calculus.
At base type we postulate the property we are trying to show, namely
existence of canonical forms.  At higher type we reduce the property to
lower types by quantifying over all possible elimination forms.

\begin{definition}[Valid Terms]
\mbox{}
\begin{enumerate}
 \item $\Psi\vd M\in\den{a}$ iff $\Psi\vd M\bua N\hastype a$, for some $N$.
 \item
  $\Psi\vd M\in\den{A\GImp{k} B}$ iff for every $\Psi'\geq \Psi$ and every
  $N$, if $ \Psi'\vd N\in\den{A}$, then $\Psi'\vd M\ N^k\in\den{B}$.
\end{enumerate}
 We say a term $M$ is \emph{valid} if $\Psi \vd M \in\den{A}$ for
appropriate $\Psi$ and $A$.
\end{definition}

First we show that all valid terms have canonical forms.  We prove at
the same time that atomic terms are valid, both by induction on the
structure of their types.

\begin{lemma}[Valid Terms have Canonical Forms] \mbox{}
\label{le:lr2canon}
\begin{enumerate}
 \item If $\Psi\vd M\in\den{A}$, then $\Psi\vd M\bua N\hastype A$.
 \item If  $\Psi\vd M\da N\hastype A$, then $\Psi\vd M\in\den{A}$.
\end{enumerate}
\end{lemma}
\begin{proof}
By induction on $A$.
\begin{description}
\item[Case] $A=a$. Immediate from the definition of $\den{a}$.
\item[Case] $A=A_1\GImp{k} A_2$.
\begin{enumerate}
\item %1
\begin{tabbing}
 $\Psi\vd M\in\den{A_1\GImp{k} A_2}$ \` Assumption \\
 $\Psi,x\oftp A_1\geq \Psi$ \` By definition of $\geq$\\
 $\Psi,x\oftp A_1\vd x\da x\hastype A_1$ \` By rule $\mathtt{tcIdvar}$\\
 $\Psi,x\oftp A_1\vd x\in\den{A_1}$ \` By i.h.\ (2)\\
 $\Psi,x\oftp A_1\vd M\ x^k\in\den{ A_2}$ \` By definition of $\den{\cdot}$\\
 $\Psi,x\oftp A_1\vd M\ x^k\bua N\hastype A_2$ \` By i.h.\ (1)\\
 $\Psi\vd M\bua\lamb{x^k}{A_1}N\hastype A_1\GImp{k} A_2$
 \` By rule $\mathtt{tc} \GImp{k} I$
\end{tabbing}
\item %2
\begin{tabbing}
 $\Psi\vd M\da M'\hastype A_1\GImp{k} A_2$ \` Assumption \\
 $\Psi'\geq\Psi$ and $\Psi'\vd N\in\den{A_1}$ for arbitrary $\Psi'$ and $N$
 \` New assumption  \\
 $\Psi'\vd N\bua N' \hastype {A_1}$ \` By i.h.\ (1) \\
 $\Psi'\vd M\da M'\hastype A_1\GImp{k} A_2$ \` By weakening \\
 $\Psi'\vd M\ N^k\da M'\ {N'}^k\hastype A_2$ \` By rule $\mathtt{tc}{\GImp{k}}E$\\
 $\Psi'\vd M\ N^k\in\den{ A_2}$ \` By i.h.\ (2)\\
 $\Psi\vd M\in\den{A_1\GImp{k} A_2}$ \` By definition of $\den{\cdot}$ 
\end{tabbing}
\end{enumerate}
\end{description}
\end{proof}

The second major part states that every well-typed term is valid.  For
this we need closure of validity under head expansion.

\begin{lemma}[Closure under Head Expansion]
\label{le:clohead}
\mbox{} \newline
 If $\Psi\vd M'\in\den{A}$ and $M\whr M'$, then $\Psi \vd M\in\den{A}$.
\end{lemma}
\begin{proof} By induction on $A$:
\begin{description}
\item[Case] $A=a$. immediate by definition and rule $\texttt{tc}\whr$.
\item[Case] $A=A_1\GImp{k} A_2$.
\begin{tabbing}
 $\Psi\vd M'\in\den{A_1\GImp{k} A_2}$ \` Assumption \\
 $\Psi'\vd N\in\den{A_1}$ for arbitrary $\Psi'\geq \Psi$ and $N$
 \` New assumption \\
 $\Psi'\vd M'\ N^k\in\den{A_2}$ \` By definition of $ \den{\cdot}$ \\
 $M\ N^k\whr M'\ N^k$ \` By rule $\nu$ \\
 $\Psi'\vd M\ N^k\in\den{A_2}$ \` By i.h.\ on $A_2$ \\
 $\Psi\vd M\in\den{A_1\GImp{k} A_2}$ \` By definition of $ \den{\cdot}$ 
\end{tabbing}
\end{description}
\end{proof}

Due to the need to $\beta$-reduce during conversion to canonical form,
we need to introduce \emph{substitutions}.  We will not require
substitutions to be well-typed, but they have to be valid in the sense
that all substitution terms should be valid.

\[ \begin{array}{rrcl}
\mbox{\textit{Substitutions}}& \theta & \bnfas \epsilon\bnfalt \theta,M/x
\end{array} \]

For $\theta=\theta',M/x$, we say that $x$ is \emph{defined} in $\theta$
and we write $\theta(x)=M$.  We require all variables defined in a
substitution to be distinct: we use $\dom(\theta)$ for the set of
variables defined in $\theta$.  Furthermore, the co-domain of $\theta$
are the variables occurring in the substituting terms.

Next, we define the \emph{application} of a substitution $\theta$ to a
term $M$, denoted $[\theta]M$.  We limit application of substitutions to
objects whose free variables are in the domain of $\theta$.
\begin{eqnarray*}
 [\theta] c & = & c \\ \relax
 [\theta] x & = & \theta(x)\\ \relax
 [\theta](M\ N^k) & = & ([\theta]M) \ ([\theta] N)^k\\ \relax
 [\theta](\lamb{x^k}{A}M) & = & \lamb{x^k}{A}[\theta,x/x]M
\end{eqnarray*}
 In the last case we assume that $x$ does not already occur in the
domain or co-domain of $\theta$.  This can always be achieved by
renaming of the bound variable.

 We will also need to mediate between single substitutions
stemming from $\beta$-reduction and simultaneous substitutions.  We
define how to \emph{compose} a single substitution from a
$\beta$-reduction with simultaneous substitutions, written
as $[N/x]\theta$.
 \begin{eqnarray*}
  [N/x](\epsilon) & = & \epsilon \\ \relax
  [N/x](\theta,M/y) & = & [N/x](\theta), ([N/x]M)/y
 \end{eqnarray*}

Note that $[N/x]([\theta,x/x]M) = [\theta,N/x]M$ if $x$ does not occur
in the co-domain of $\theta$.  For a context $\Psi=x_1\oftp A_1,\ldots,
x_n\oftp A_n$, we introduce the \emph{identity} substitution on $\Psi$
as ${\rm id}_{\Psi}=x_1/x_1,\ldots, x_n/x_n$.  Clearly, ${\rm id}_{\Psi}M
= M$ if the free variables of $M$ are contained in $\Psi$.

We extend the notion of validity to substitutions as already indicated
above: a substitution $\theta$ is valid for context $\Psi$ if for
every binding $M/x$ such that $x\oftp A$ is in $\Psi$ we have $M$ is in
$\den{A}$.

\begin{definition}[Valid Substitutions]
\mbox{}
\begin{enumerate}
 \item $\Phi\vd \theta\in\den{\cdot}$ iff $\theta=\epsilon$.
 \item $\Phi\vd\theta\in\den{\Psi',x\oftp A}$
   iff $\theta=\theta',M/x$ such that $\Phi\vd M\in\den{A}$ and
   $\Phi\vd\theta'\in\den{\Psi'}$.
\end{enumerate}
\end{definition}

We remark that contexts are not ordered, hence, for
$\Psi=(\GG,\OO,\DD)$ we will identify, for example, 
$\den{\Psi,x\oftp A}$ with $\den{(\GG,x\oftp A,\OO,\DD)}$.  Clearly, this
view is legitimate in terms of the above definition, since validity
of a substitution simply reduces to validity of the terms in
it.  It is easy to see that validity, both
for terms and for substitutions, satisfies weakening.  We omit
the formal statement and proof of this property.

The next lemma is critical.  It generalizes the statement that
well-typed terms are valid by allowing for a valid substitution
to be applied.  This is necessary in order to proceed with the
proof in the case of any of the three $\lambda$-abstractions.

\begin{lemma}[Well-Typed Terms are Valid]
\label{le:lrmain} \mbox{} \newline
 If $\god\vd M\hastype A$, then for every $\Psi$ such that
$\Psi\vd\theta\in\den{\godc}$ we have $\Psi\vd [\theta]M\in\den{A}$.
\end{lemma} 
\begin{proof}
 By induction on the typing derivation $\cD$ of $\god\vd M\hastype A$.
\begin{description}
\item[Case]
 \[ \cD= \ianc{}{(\GG, x\oftp A); \OO; \cdot \vd x \hastype A}{\mathtt{Id}^u} \]
\begin{tabbing}
 $\Psi\vd\theta\in\den{(\GG,x\oftp A,\OO)}$ \` Assumption \\
 $\Psi\vd\theta(x)\in\den{A}$ \` By definition of $\den{\cdot}$\\
 $\Psi\vd[\theta]x\in\den{A}$ \` By definition of substitution
\end{tabbing}
\item[Case] $\cD$ ends in $\mathtt{Id}^1$.  As in the previous case.
\item[Case] $\cD$ ends in $\mathtt{Con}$. Immediate by
Lemma~\ref{le:lr2canon}(2) and definition of substitution.
\item[Case] % TImp
 \[ \cD=\ianc{(\GG, x\oftp A); \OO; \DD \vd M \hastype B}{\GG;
\OO; \DD \vd \lam x^u \oftp A\ldot M \hastype A \GImp{u}
B}{\GImp{u}I} \]
\begin{tabbing}
 $(\GG, x\oftp A); \OO; \DD \vd M \hastype B$ \` Subderivation \\
 $\Psi\vd\theta\in\den{\godc}$ \` Assumption \\
 $\Psi'\vd N\in\den{A}$ for arbitrary $\Psi' \geq \Psi$ and $N$
 \` New assumption \\
 $\Psi'\vd(\theta,N/x)\in\den{(\GG,x\oftp A,\OO,\DD)}$
 \` By definition of $\den{\cdot}$
 and weakening\\
 $\Psi'\vd [\theta,N/x]M\in\den{B}$ \` By i.h. \\
 $\Psi'\vd [N/x]([\theta,x/x] M)\in\den{B}$ \` By property of substitution \\
 $\Psi'\vd(\lamb{x^u}{A}[\theta,x/x]M) N^u\in\den{B}$ \` By Lemma~\ref{le:clohead} \\
 $\Psi'\vd ([\theta](\lamb{x^u}{A}M)) N^u\in\den{B}$
 \` By definition of substitution \\
 $\Psi\vd [\theta](\lamb{x^u}{A}M)\in\den{A\TImp B}$
 \` By definition of $\den{A\TImp B}$
\end{tabbing}
\item[Cases] $\cD$ ends in ${\GImp{0}I}$ or ${\GImp{1}I}$.  Analogous
to previous case.
\item[Case]
 \[ \cD = \ibnc{\GG; \OO; \DD \vd M \hastype A \TImp B}{(\GG,
\DD); \OO; \cdot \vd N \hastype A}{\GG; \OO; \DD \vd M\
N^u \hastype B}{\GImp{u}E} \]
\begin{tabbing}
 $\Psi\vd\theta\in\den{\godc}$ \` Assumption \\
 $\GG; \OO; \DD \vd M \hastype A \TImp B$ \` Subderivation \\
 $\Psi\vd[\theta]M\in\den{A\TImp B}$ \` By i.h. \\
 $(\GG,\DD); \OO; \cdot \vd N \hastype A$ \` Subderivation \\
 $\Psi \vd [\theta]N \in\den{ A}$ \` By i.h. \\
 $\Psi\geq\Psi$ \` By definition of $\geq$\\
 $\Psi\vd ([\theta]M)([\theta] N)^u\in\den{B}$ \` By definition
 of $\den{\cdot}$ \\ 
 $\Psi\vd [\theta](M\ N)^u\in\den{B}$ \` By definition of substitution
\end{tabbing}
\item[Cases] $\cD$ ends in ${\GImp{0}E}$ or ${\GImp{1}E}$.  Analogous
to the previous case.
\end{description}
\end{proof}

From this central lemma, the canonical form theorem follows by
noting that the identity substitution is valid.

\begin{lemma}[Validity of Identity]
\label{le:lrid}
 $\Psi\vd id_{\Psi}\in\den{\Psi}$.
\end{lemma}
\begin{proof}
  By a straightforward induction on $\Psi$ using
  Lemma~\ref{le:lr2canon}(2).
\end{proof}

\begin{theorem}[Canonical Forms]
\label{thm:canonical} \mbox{} \newline
 If $\god\vd M\hastype A$, then there exists an $N$ such that $\godc\vd
  M\bua N\hastype A$ and $\god\vd N\bua A$.
\end{theorem}
\begin{proof} Direct from prior lemmas.
\begin{tabbing}
 $\god\vd M\hastype A$ \` Assumption \\
 $\godc\vd id_{\godc}\in\den{\godc}$ \` By Lemma~\ref{le:lrid} \\ 
 $\godc\vd [id_{\godc}]M\in\den{A}$ \` By Lemma~\ref{le:lrmain} \\ 
 $\godc\vd M\in\den{A}$ \` By identity substitution \\
 $\godc\vd M\bua N\hastype A$ for some $N$ \` By Lemma~\ref{le:lr2canon}(1) \\
 $\god\vd  N\bua A$ \` By Theorem~\ref{thm:convercf}
\end{tabbing}
\end{proof}

We close this section with some remarks on related work on strictness.
Church original definition of the set $\Lambda_{I}$ of (untyped)
$\lam$-terms \cite{Church41} has this clause for abstraction:
\begin{center}
 If $M\in\Lambda_{I}$ and $x\in FV(M)$, then $\lam x\ldot M\in\Lambda_{I}$.
\end{center}
 Therefore, in this language there cannot be any vacuous abstractions.
The combinatorial counterpart of this calculus excludes $\mathbf{K}$ and
consists of $\mathbf{I,W,B,C}$.  Those are the axioms of what Church
called \emph{weak implicational logic}~\cite{Church51}, i.e., identity,
contraction, prefixing and permutation.  This establishes the link with
an enterprise born from a very different origin, namely the relevance
logic project~\cite{Anderson75}, which emerged in fact in the early
sixties out of Anderson and Belnap's dissatisfaction with the so-called
``\emph{paradoxes of implication}'', let it be material, intuitionistic,
or strict (in the modal sense of Lewis and Langford).

 Following Girard's and Belnap's suggestion \cite{Belnap93}, we will
\emph{not} refer to our calculus as \emph{relevant}, but as
\emph{strict} logic, as the former may also satisfy other principles
such as distributivity of implication over conjunction.

 On an unrelated front, starting with Mycroft's seminal paper
\cite{Mycroft80}, compile-time analysis of functional programs
concentrated on strictness analysis in order to get the best out of
call-by-value and call-by-need evaluation; first in terms of abstract
interpretation, later by using non-standard types to represent these
``intensional'' properties of functions (see \cite{Jensen91} for a
comparison of these two techniques).  However, earlier work such as
\cite{Kuo89} used non-standard primitive type to distinguish strict or
non-strict terms, closed only under unrestricted function space.
In the setting of functional programming, various different notions
of strictness emerged.  However, the absence of recursion and
effects in our setting admits fewer distinctions.

\citeN{Wright92} seems to be the first to have extended the
Curry-Howard isomorphism to the implicational fragment of relevance
logic and explicitly connected the two areas, although both
\cite{Belnap74} and \cite{Helman77} had previously recognized the link
between strictness and relevance.

\citeN{Baker94} presents a type assignment system that
makes available strict, invariant and intuitionistic types.  It is
biased towards enforcing strictness information, which ultimately leads
to a different expressive power from our calculus.  There is only one
context, where variables carry their occurrence status as a label.
There is one identity rule, the strict one, so that e.g.\ $\lam x\ldot x
\hastype A\TImp A$ is not derivable, as it can be given the more stringent type
$A\RImp A$.  Let us consider the elimination rules for strict and
irrelevant functions.
 \[ \begin{array}{c}
 \ibnc{\GG\vd M\hastype A\RImp B}{\GG'\vd N\hastype A'}{\GG,\GG'\vd
   M\ N\hastype B}{app\RImp}\\[1ex]
 \ibnc{\GG\vd M\hastype A\ZImp
   B}{\GG'\vd N\hastype A'}{\GG,\GG'[1:=0]\vd M\ N\hastype
   B}{app\ZImp}
   \end{array} \]
 A side condition $A'\leq A$ enforces the information ordering, so that
for example $A'\ZImp B\leq A\TImp B'$, provided that $A\leq A', B\leq
B'$.  This allows us to infer by strict application $\GG,\GG'\vd M\
N\hastype C$ from $\GG\vd M\hastype (A\TImp B)\RImp C$ and $\GG'\vd N
\hastype A\ZImp B$. The latter is instead forbidden in our system by the
labeled reduction rules.  The rationale on the relabeling operation in
the rule $app\ZImp$ is that $A$ is not relevant to $B$, so all
hypothesis should be deleted.  Instead, in order to preserve every
variable declaration, their strict label is changed into irrelevant.
This would amount to moving the strict variables in the irrelevant
context in our system. Note the difference with our rule, where the
latter variables are moved in the unrestricted context.  Moreover,
having only one context, the author needs a strategy to deal with
the same variable with different annotations; the solution is that while
propagating premises top-down a binding $x^1\oftp A$ supersedes
$x^u\oftp A$ which in turn supersedes $x^0\oftp A$.
 
\citeN{Wright96} introduces \emph{Annotation Logic} as a
general framework for resource-conscious logics. Its formulae have the
form $A\bnfas X^k\bnfalt A\GImp{k} B$ for any annotation $k$ and there
are specific structural as well as annotation rules.  The latter
implement rules such as promotion or dereliction.  By instantiation with
different algebras of annotation, we get systems such as linear and
strict logic as well as various other usage logics.  An abstract
normalization procedure is sketched, which however requires commutative
conversions already in the purely implicational fragment.

In summary, none of the systems of strict function in the literature
served our purpose, nor did any of the authors prove the existence
of canonical forms that are critical for our application.

\section{Simple Terms}
\label{sec:strictcomp}

Now that we have developed a calculus which is potentially strong enough
to represent the complement of linear patterns, two questions naturally
arise: how do we embed the original $\lambda$-calculus, and is the
calculus now closed under complement?  We require that our complement
operator ought to satisfy the usual boolean rules for negation:
\begin{enumerate}
 \item (Exclusivity) It is not the case that some $M$ is both a ground
instance of $N$  and of $\mnot(N)$.
 \item (Exhaustivity) Every $M$ is a ground instance of $N$ or of $\mnot(N)$.
\end{enumerate}
 Remember that when we refer to \emph{ground instances} we mean instances
without any existential variables.  Parameters, on the other hand, can
certainly occur.

Unfortunately, while the first property follows quite easily for a
suitable algorithm, it turns out the second cannot be achieved for the
full strict $\lambda$-calculus calculus as presented in the previous
sections.  The following counterexample is a pattern whose complement
cannot be expressed within the language.

\begin{example}
 Consider the signature $a\oftp\mbox{type}, b\oftp a, c\oftp a \TImp a$.
Then in the context $x\oftp a; \cdot; \cdot$ we have
 \[ \begin{array}{rcl}
  \| E\ x^0 \| & = & \{ b, c\ b^u, c\ (c\ b^u)^u, \ldots \} \\
  \mnot(\| E\ x^0 \|) & = & \{ x, c\ x^u, c\ (c\ x^u)^u, \ldots \} \\
 %  \| F\ x^1 \| & = & \{ x \} \\
 % \mnot(\| F\ x^1 \|) & = & \{ c\ x^u, c\ (c\ x^u)^u, \ldots, b, c\ b^u, \ldots \}
 \end{array} \]
 It is easy to see that $\mnot(\| E\ x^0\|)$ cannot be described by a
finite set of patterns.  The underlying problem is the undetermined
status of the argument to $c \oftp a \TImp a$ which means it can 
contain neither strict nor irrelevant variables while being allowed to
contain unrestricted variables.
 \end{example}

However, the main result of this section is that the complement
algorithm presented in Definition~\ref{def:compl} is sound and complete
for the fragment which results from the natural embedding of the
original simply-typed $\lambda$-calculus; this is sufficient for our
intended applications.  We will proceed in two phases.  First we
restrict ourselves to a class of terms (that we call \emph{simple}) for
which the crucial property of \emph{tightening} (Lemma~\ref{le:exh}) can
be established.  Second we transform the complement problem so that each
existential variable is applied to \emph{all} parameters and bound
variables in whose scope it appears.  This improvement is mainly
cosmetic and makes it easier to state and prove correctness for our
algorithms.

Recall that we have introduced strictness to capture occurrence
conditions on variables in canonical forms.  This means that first-order
constants (and by extension bound variables) should be considered
\emph{strict} functions of their argument, since these arguments will
indeed occur in the canonical form.  On the other hand, if we have a
second order constant, we cannot restrict the argument function to be
either strict or vacuous, since this would render our representation
language too weak.

\begin{example}
\label{ex:excont}
 Continuing Example~\ref{ex:lam}, consider the representation of the
\textbf{K} combinator:
\begin{eqnarray*}
 \rep{\Lambda x\ldot \Lambda y\ldot x}& = & \itlam\ (\lamb{x}{\itexp} \itlam\ (\lamb{y}{\itexp} x))
\end{eqnarray*}
Notice that the argument to the first occurrence of `$\itlam$' is a
strict function, while the argument to the second occurrence is an
invariant function.  If we can give only one type to `$\itlam$' it must
therefore be $(\itexp \GImp{u} \itexp) \GImp{1} \itexp$.
\end{example}

Generalizing this observation means that positive occurrence of
function types are translated to strict functions, while the negative
ones to undetermined functions.  We can formalize this as an
embedding of the simply-typed $\lam$-calculus into a fragment of the
strict calculus via two (overloaded) mutually recursive translations
$()^{-}$ and $()^{+}$.  First, the definition on types:
\begin{eqnarray*}
 (A\arrow B)^{+} & = & A^{-} \RImp B^{+}\\
 (A\arrow B)^{-} & = & A^{+} \TImp B^{-}\\
 a^{-} = a^{+} & = & a 
\end{eqnarray*} 
 We extend it to atomic and canonical terms (including existential
variables), signatures, and contexts; we therefore need the usual
inductive definition of atomic and canonical terms in the simply-typed
$\lam$-calculus (see for example~\cite{Pfenning01book}), which can be
obtained by dropping labels from the definition of canonical form in
Figure~\ref{fig:canon}.  In addition, we allow well-typed applications
$E_A\ x_1^{k_1}\ldots x_n^{k_n}$ of base type as canonical terms.
Recall that $x_1, \ldots, x_n$ must be distinct bound variables or
parameters.  Note that the embedding $()^{-}$ is applied only to
canonical terms, while $()^{+}$ is applied only to atomic terms.

\begin{eqnarray*}
 (\lam x\oftp A\ldot M)^{-} & = & \lamb{x^{u}}{A^{+}} M^{-} \\
 (E_A\ x_1\ldots x_n)^{-} & = & F_{A^{-}} \ x_1^u\ldots x^u_n \\
 M^{-} & = & M^{+} \qquad\qquad \mbox{for $M$ of base type} \\
 x^{+} & = & x\\
 c^{+} & = & c\\
 (M\ N)^{+} & = & M^{+}\ (N^{-})^{1} \\[1em]
 (\cdot)^{+} & = & \cdot \\
 (\GG,x\oftp A)^{+} & = & \GG^{+}, x\oftp A^{+}\\
 (\Sigma,a\oftp type)^{+} & = & \Sigma^{+}, a\oftp type\\
 (\Sigma,c\oftp A)^{+} & = & \Sigma^{+}, c\oftp A^{+} \end{eqnarray*}

\begin{example} Returning to Example~\ref{ex:excont}:
\begin{eqnarray*}
 (\itlam\ (\lamb{x}{\itexp} \itlam\ (\lamb{y}{\itexp} x)))^+& = & 
 \itlam\ (\lamb{x^u}{\itexp} \itlam\ (\lamb{y^u}{\itexp} x)^1)^1
\end{eqnarray*}
\end{example}

The image of the embedding of the canonical forms of the simply-typed
$\lambda$-calculus gives rise to the following fragment, where we
allow existential variables to have arguments with arbitrary
labels.
\[ \begin{array}{rrcl}
 \mbox{\textit{Simple Terms}} & M & \bnfas &  \lam x^{u}\oftp
A^{+}\ldot M \bnfalt h\ M_1^{1} \ldots M_n^{1} \bnfalt 
 E_{A}\ x_{1}^{k_1}\ldots x_{n}^{k_n}
\end{array} \]
 It is possible to generalize this language further to allow
arbitrary abstractions as well, but this is beyond the scope
of the present paper (see the comment in the Section~\ref{sec:concl}).

\begin{theorem}[Correctness of $()^{\pm}$]
\label{thm:embed}
\mbox{}
\begin{enumerate}
 \item If $\GG\vd M\bua A$, then $\GG^{+};\cdot;\cdot \vd M^{-}\bua A^{-}$. 
 \item If $\GG\vd M\da A$, then $\GG^{+};\cdot;\cdot \vd M^{+}\da A^{+}$. 
\end{enumerate}
\end{theorem}

\begin{proof}
 By mutual induction on the structure of the derivations of $\GG\vd
M\bua A$ and $\GG\vd M\da A$.
\end{proof}

From now on we may hide the $()^1$ decoration from strict application of
constants in examples. Moreover, we will shorten judgment $\cal J$ on
simple terms of the form $\Psi; \cdot; \cdot \vd {\cal J}$ to $\Psi \vd
{\cal J}$.

We can now prove the crucial tightening lemma.  It expresses the
property that every simple term with no existential variable is either
strict or vacuous in a given undetermined variable.

\begin{lemma}[Tightening]
\label{le:exh} 
Let $M$ be a simple term of type $A$ with no existential variables.
\begin{enumerate}
 \item If $(\GG, x\oftp C); \OO; \DD \vd M\da A$ then \newline
  either $\GG; \OO; (\DD, x\oftp C) \vd M\da A$ or $\GG; (\OO, x\oftp
C); \DD \vd M\da A$.
 \item If $(\GG, x\oftp C); \OO; \DD \vd M\bua A$ then \newline
  either $\GG\; \OO; (\DD, x\oftp C) \vd M\bua A$ or $\GG; (\OO, x\oftp
C); \DD \vd M\bua A$.
\end{enumerate}
\end{lemma}
\begin{proof}
By mutual induction on $\cD_{1}::(\GG, x\oftp C); \OO; \DD \vd M\da A$ and 
$\cD_{2}::(\GG, x\oftp C); \OO; \DD \vd M\bua A$.  We show only one case.
\begin{description}
\item[Case]
 \[ \cD_1 = \ianc{(\GG,x\oftp C,\DD_N);\OO;\DD_M \vd M \da A \RImp B
\hspace{1em} (\GG,x\oftp C,\DD_M);\OO;\DD_N \vd N \bua A}{(\GG,x\oftp
C);\OO; (\DD_{M},\DD_{N}) \vd M\ N^1 \da B}{\mathtt{c}{\RImp}E} \]
 There are four sub-cases, stemming from the two possibilities
each for the two subderivations.
\begin{enumerate}
\item
\begin{tabbing}
 $(\GG,\DD_N);\OO;(\DD_M,x\oftp C)\vd M\da A\RImp B$ \` Subcase of i.h. \\
 $(\GG,\DD_{M});\OO;(\DD_N,x\oftp C)\vd N\bua A$ \` Subcase of i.h.  \\
 $(\GG,\DD_M,x\oftp C);\OO;\DD_N\vd N\bua A$ \` By Loosening$^{1}$ $x$\\
 $\GG;\OO;(\DD_M,x\oftp C,\DD_N)\vd M\ N^{1}\da B$ \` By rule $\texttt{c}{\RImp}E$
\end{tabbing}
\item 
\begin{tabbing}
 $(\GG,\DD_N);(\OO,x\oftp C);\DD_M\vd M\da A\RImp B$ \` Subcase of i.h. \\
 $(\GG,\DD_M);(\OO,x\oftp C);\DD_N\vd N\bua A$ \` Subcase of i.h. \\
 $\GG;(\OO,x\oftp C);(\DD_{M},\DD_{N})\vd M\ N^{1}\da B$ \` By rule  $\texttt{c}{\RImp}E$
\end{tabbing}
\item
\begin{tabbing}
 $(\GG,\DD_N);\OO;(\DD_M,x\oftp C)\vd M\da A\RImp B$ \` Subcase of i.h. \\
 $(\GG,\DD_M);(\OO,x\oftp C);\DD_N\vd N\bua A$ \` Subcase of i.h. \\
 $(\GG,\DD_M,x\oftp C);\OO;\DD_N\vd N\bua A$ \` By Loosening$^{0}$ $x$ \\
 $\GG;\OO;(\DD_M,x\oftp C,\DD_N)\vd M\ N^{1}\da B$ \` By rule  $\texttt{c}{\RImp}E$
\end{tabbing}
\item Symmetrical to (3).
\end{enumerate}
\end{description}
\end{proof}

We remark that tightening fails to hold once we allow unrestricted function
types in a negative position.  For example, $(y\oftp A\TImp B,x\oftp
A);\cdot;\cdot \vd y\ x^u\hastype B$ but both $y\oftp A\TImp B;\cdot;x\oftp
A\nvd y\ x^u\hastype B$ and $y\oftp A\TImp B;x\oftp A;\cdot\nvd y\ x^u\hastype
B$.

We also have the following related property.

\begin{lemma}[Irrelevance]
\label{le:fat}
 Let $M$ be a simple term without existential variables.
\begin{enumerate}
 \item If $\GG; (\OO, x\oftp C); \DD \vd M\bua A$,
   then $\GG; \OO; \DD \vd M\bua A$.
 \item If $\GG; (\OO, x\oftp C); \DD \vd M\da A$,
  then $\GG; \OO; \DD \vd M\da A$.
\end{enumerate}
\end{lemma}

\begin{proof}
By mutual induction on the given derivations.
\end{proof}

Note that irrelevance holds for any strict canonical term, but it is
false for terms containing redices.  For example, for
$c\oftp B$ we have $\cdot;x\oftp A;\cdot\vd
(\lamb{y^0}{A} c)\ x^0\hastype B$, but $\cdot;\cdot;\cdot\nvd
(\lamb{y^0}{A} c)\ x^0\hastype B$.

For simple terms it is often more convenient to replace explicit reference
to atomic forms by an $n$-ary version of $\mathtt{c}{\RImp}E$.  This can
easily be seen to cover all atomic forms for simple terms where
the head $h$ can be a variable $x$ or constant $c$.
 \[ \ianc{\Psi\vd h\hastype A_1\RImp\cdots\RImp A_n\RImp a \quad
       \Psi \vd N_1\bua A_1 \quad \cdots \quad \Psi \vd N_n \bua A_n}
{\Psi \vd h\ N_1^1\ldots N_{n}^1\bua a}{\texttt{c}{\RImp}E} \]
      
We can simplify the presentation of the algorithms for complement and
later unification if we require any existential variable to be applied
to every bound variable in its declaration context.  This is possible
for any simple linear pattern without changing the set of its
ground instances.  We just insert vacuous applications, which
guarantee that the extra variables are not used. 

In a slight abuse of notation we call the resulting patterns \emph{fully
applied}.  This transformation is entirely straightforward and its
correctness is easily established using Irrelevance
(Lemma~\ref{le:fat}). We omit the formal details here, showing only an
example.

\begin{example}
  Recall the simple pattern that encodes an object-level $\eta$-redex
from Example~\ref{ex:etardx},
 \[ \itlam\ (\lamb{x^u}{\itexp} \itapp\ E\ x). \]
  It is not fully applied, since $E$ is not applied to $x$.  This is crucial,
since $E$ is not allowed to depend on the bound variable $x$.
In its fully applied form
 \[ \itlam\ (\lamb{x^u}{\itexp} \itapp\ (E'\ x^0)\ x), \]
 this occurrence condition is encoded by an irrelevant application of a
fresh existential variable $E'$ of type $\itexp \GImp{0} \itexp$ to $x$.
According to Lemma~\ref{le:fat}, this means that $x$ cannot occur in the
canonical form of $E'\; x^0$ for any instance of $E'$.
\end{example}

In the remainder of this paper we will assume that all existential variables
are fully applied as defined above.  We refer to a pattern $E\, x_1^{k_1}\ldots
x_n^{k_n}$ as a \emph{generalized variable}.  Furthermore, we always sort the
variables $x_1\ldots x_n$ so that they come in some standard order; this
simplifies the description of some of the algorithms on fully applied
patterns.  Following standard terminology we call atomic terms whose head is
a bound variable or a parameter \emph{rigid}, while terms whose head is an
existential variable is called \emph{flexible}.

Under these assumptions we can more formally specify the interpretation
of terms with existential variables.  We use $\Phi$ for sequences of
distinct, labelled bound variables; if $x^k\in\Phi$, we set $\Phi(x)=
k$. We say that $\god\vd\Phi\ ok$ if the following holds:
\begin{eqnarray*}
 \Phi(x) = u & \BImp & x\in\dom(\GG)\\
 \Phi(x) = 0 & \BImp & x\in\dom(\OO)\\
 \Phi(x) = 1 & \BImp & x\in\dom(\DD)
\end{eqnarray*}
Note that $\Phi$ determines $\GG; \OO; \DD$ and vice versa whenever
$\god\vd \Phi\ ok$.

\begin{figure}
\[
\begin{array}{c}
 \ianc{\GG; \OO; \DD \vd \Phi\; ok \qquad
       \GG; \OO; \DD \vd M \hastype a}
      {\Psi \vd M \in\| E_A\ \Phi\|\hastype a}
      {\mbox{\tt grFlx}}
\\[2ex]
\ianc{(\Psi,x\oftp A)\vd   M  \in\| N\|\hastype B}
{\Psi \vd  \lam x^u\oftp A\ldot M  \in\|\lam x^u\oftp A\ldot N \|:A\TImp B}{\mbox{\tt grLam}}
\\[2ex]
\ianc{\Psi \vd h \hastype  A_1\RImp\cdots\RImp A_n\RImp a \qquad
      \Psi \vd M_1 \in\| N_{1}\|\hastype A_1\;\cdots\;
      \Psi\vd M_n\in\|N_n\| \hastype A_n}
     {\Psi \vd  h\ M_1^1\ldots M_n^1 \in\|h\ N_1^1\ldots N_n^1 \|\hastype a}
     {\mbox{\tt grApp}}
\end{array}
\]
\caption{Ground instance: $\Psi \vd M\in  \|N\| \hastype A$}
\label{fig:grinst}
\end{figure}

Recall that every pattern can be seen as the intensional
representation of the set of its instances with respected to a fixed
signature $\Sigma$ and a set of parameters declared in a context
$\Psi$.  The judgment in Figure~\ref{fig:grinst}, $\Psi \vd M\in\|
N\|\hastype A $, formalizes the conditions for $M$ canonical of type
$A$ to be a ground instance of a simple \emph{linear} pattern $N$ at
type $A$.

\begin{remark}
\label{rmrk:gr}
Note that $\Psi \vd M \in \|E_A\ \Phi\| \hastype a$ means that $M$ is
indeed a ground instance of $E_A\ \Phi$. Conversely, if $\Phi =
x_1^{k_1} \ldots x_n^{k_n}$ and $A = A_1 \GImp{k^1} \cdots 
A_n \GImp{k^n} a$ then we set $E_A = \lam x_1^{k_1}\oftp A_1\ldots \lam
x_n^{k_n}\oftp A_n\ldot M$
\end{remark}

\section{The Complement Algorithm}
\label{sec:compl}

The idea of complementation for atomic terms and abstractions is quite
simple and similar to the first-order case.  For generalized variables
we consider each argument in turn.  If an argument variable is
undetermined it does not contribute to the negation.  If an argument
variable is strict, then any term where this variable does not occur
contributes to the negation.  We therefore complement the
corresponding label from $0$ to $1$ while all other arguments are
undetermined.  For vacuous argument variables we proceed dually.  

In preparation for the rules, we observe that the complement operation
on patterns behaves on labels like negation does on truth-values in
Kleene's three-valued logic, in the sense of the following table:.
 \[ \begin{array}{c}
 \mnot(1) = 0 \qquad
 \mnot(0) = 1 \qquad
 \mnot(u) = u
 \end{array} \]

We extend this definition to sequences of variables as they are used to
codify occurrence constraints for existential variables.
\begin{eqnarray*}
 \mnot_i(x_{1}^{k_{1}}\ldots
 x_{i-1}^{k_{i-1}} \ x_{i}^{d}\ x_{i+1}^{k_{i+1}}\ldots x_n^{k_{n}})
 & = &
 x_{1}^u\ldots x_{i-1}^u \ 
 x_{i}^{\mnot(d)}\ x_{i+1}^u\ldots x_n^u
\end{eqnarray*}
Note that we require $x_i$ to be determined ($d \in \{0,1\}$) for
$\mnot_i$ to be defined, and that variables $x_j$ for $j \not= i$ are
all unrestricted on the right-hand side even though their status on
the left-hand side varies.

\begin{figure}
\[
\begin{array}{c}
\ianc{\mbox{$\mnot_i(\Phi)$ defined}}
     {\Psi\vd \mnot(E \ \Phi) \blip  Z\ \mnot_i(\Phi) \hastype a}
     {\mathtt{NotFlx}^i}
\\[2ex]
\ianc{\Psi,x\oftp A\vd  \mnot(M)  \blip   N\hastype B}
{\Psi\vd  \mnot(\lam x^u\oftp A\ldot M)  \blip  \lam x^u\oftp A\ldot N\hastype A\TImp B}{\mathtt{NotLam}}
\\[2ex]
 \ianc{g\in\mathrm{dom}(\Sigma\cup\Psi), g \hastype A_{1}\RImp\ldots\RImp
A_{m}\RImp a, h \not= g}
 {\Psi\vd  \mnot (h\ M_1^1\ldots M_n^1) \blip  
g\ (Z_{1} \ \Psi^{u})^{1}\ldots (Z_{m}\ \Psi^{u})^{1}\hastype a}
 {\mathtt{NotApp}_{1}}
\\[2ex]
\ianc{\Psi\vd \mnot( M_i) \blip  N\hastype A_i}
{\Psi\vd  \mnot (h\ M_1^1\ldots M_n^1) \blip  
h\ (Z_{1}\ \Psi^{u})^{1}\ldots (Z_{i-1}\ \Psi^{u})^{1} \ N^{1}\ 
(Z_{i+1}\ \Psi^{u})^{1}\ldots (Z_{n}\ \Psi^{u})^{1}\hastype a}
{\mathtt{NotApp}_{2}^i}
\end{array}
\]
\caption{Complement algorithm: $\Psi\vd \mnot(M) \blip N\hastype A$}
  \label{fig:compl}
\end{figure}

\begin{definition}[Higher-Order Pattern Complement]
\label{def:compl}
For a linear simple pattern $M$ such that $\Psi \vd M \Uparrow A$, define $\Psi
\vd \mnot(M) \blip N\hastype A$ by the rules in Figure~\ref{fig:compl}, where
the $Z$'s are fresh logic variables of appropriate type,
$h\in\mathrm{dom}(\Sigma\cup\Psi)$ and $\Psi\vd h\hastype A_{1}\RImp\ldots\RImp
A_{n}\RImp a$.  We write $Z\ \Psi^u$ as an abbreviation for $Z\ \Phi$ where
$\Psi; \cdot; \cdot \vd \Phi\ ok$.

Note that a given $M$ may be related to several patterns $N$ all of which
belong to the complement of $M$.  We therefore define $\Psi\vd\mnot(M)=
\cN\hastype A$ if $\cN=\{N\mid\Psi\vd \mnot(M)\blip N\hastype A\}$.
\end{definition}

We may drop the type information from the above judgment in examples
and proofs; we will write $\Psi\vd M \in \| \mnot(N) \|\hastype A$, when
$\Psi\vd\mnot(N)=\cN$ and $\Psi\vd M \in \| \cN \|\hastype A$. 

\begin{example} Consider the following complement problems.
\begin{eqnarray}
 x\oftp a, y\oftp a \vd \mnot (E\ x^{u}\ y^{1}) & = & \{ F\ x^{u}\ y^{0} \}\nonumber \\
 x\oftp a, y\oftp a \vd \mnot(E\ x^{0}\ y^{1}) & = & \{ F\ x^{1}\ y^{u}, G\ x^{u}\ y^{0} \} 
\label{eq:l}
\end{eqnarray}
\end{example}

It is worthwhile to observe that the members of a complement set are not
mutually disjoint, due to the indeterminacy of $u$.  We can achieve
exclusive patterns if we resolve this indeterminacy by considering for
every $x^{u}$ the two possibilities $x^{1},x^{0}$.  Thus, for example,
the right-hand side of equation (\ref{eq:l}) can be rewritten as
 \[ \{F \ x^{1} y^{1},    G \ x^{1} y^{0},  H \ x^{0} y^{0}\}. \]
It is clear that in the worst case scenario the number of patterns in a 
complement set is bounded by $2^{n}$; hence the usefulness of this 
further step needs to be pragmatically determined.

We can now revisit the example of an $\eta$-redex in the untyped
$\lambda$-calculus.  To avoid too many indices on existential variables,
we adopt a convention that the scope of existential variables is limited
to each member of a complement set.

\begin{example}
 Reconsider Example~\ref{ex:etardx}.  Then we calculate:
\[ \begin{array}{rcl}
 \cdot & \vd & \mnot(\itlam(\lamb{x^{u}}{\itexp} \itapp\ (E\ x^{0})\ x))\\
& & = \{
 \begin{array}[t]{@{}l}
  \itlam(\lamb{x^u}{\itexp} \itapp\ (Z\ x^1)\ (Z'\ x^u)),\\
  \itlam(\lamb{x^u}{\itexp} \itapp\ (Z\ x^u)\ (\itapp\ (Z'\ x^u) \ (Z''\  x^u)), \\
  \itlam(\lamb{x^u}{\itexp} \itapp\ (Z\ x^u)\ ( \itlam(\lamb{y^u}{\itexp} Z'\ x^u\ y^u)),\\
  \itlam(\lamb{x^u}{\itexp} \itlam(\lamb{y^u}{\itexp} Z\ x^u\ y^u) ),\\
  \itlam(\lamb{x^u}{\itexp}x),\\
  \itapp\ Z\ Z'\}
  \end{array}
\end{array} \]
\end{example}

We now address the correctness of the complement algorithm with respect to the
set-theoretic semantics.  The proof obligation consists in proving that the
former does behave as a complement operation on sets of patterns, that is, it
satisfies disjointness and exhaustivity.  Disjointness is the property that a
set and its complement share no element; exhaustivity states that every
element is in the set or its complement.  Termination is obvious as the
algorithm is syntax-directed and only finitely branching.  We start with
disjointness between a pattern and its complement.

\begin{lemma}[Disjointness of Complementation]
\label{le:part1}
\mbox{} \newline
Let $\Psi\vd N\bua A$ be a simple linear pattern.  Then for every $Q$
such that $\Psi\vd \mnot(N)\blip Q\hastype A$, it is not the case that
both $\Psi\vd M \in \| N\|\hastype A$ and $\Psi\vd M \in \| Q\|\hastype
A$.
\end{lemma}
\begin{proof}
  By induction on the structure of $\cD::\Psi\vd \mnot(N)\blip Q\hastype A$.
 \begin{description}
 \item[Case]
  $\cD$ ends in $\mnot\mbox{\tt Flx}^i$.
 \begin{tabbing}
  $\Psi \vd M \in \|E\ \Phi\|\hastype a$ \` Assumption \\
  $\Psi \vd M \in \|Z\ \mnot_i(\Phi)\|\hastype a$ \` Assumption \\[0.5ex]
  $\Phi(x_i) = 1$ or $\Phi(x_i) = 0$ \` Since $\mnot_i(\Phi)$ defined
 % $\Phi(x_i) = 1$ \` Subcase \\
\end{tabbing}
\begin{description}
\item[Subcase]  $\Phi(x_i) = 1$
 \begin{tabbing}
  $\GG; \OO; (\DD, x_i\oftp A) \vd M \hastype a$ \` By
inversion on $M \in \|E\ \Phi\|$ \\
   $(\GG, \OO, \DD); x_i\oftp A; \cdot \vd M \hastype a$
\` By inversion on $M \in \|Z\ \mnot_i(\Phi)\|$ \\
   $\perp$ \` By exclusivity (Lemma~\ref{le:disj})
\end{tabbing}
%\begin{description}
\item[Subcase]  $\Phi(x_i) = 0$ is symmetrical.
 % \begin{tabbing}
 %  $\Phi(x_i) = 0$ \` Subcase \\
 % $\perp$ \` Symmetrical 
 % \end{tabbing}
\end{description}
% \end{tabbing}

 \item[Case] $\cD$ ends in $\mathtt{NotApp}_{1}$.
 \begin{tabbing}
  $\Psi \vd M \in \|h\ N_1^1\ldots N_n^1\|\hastype a$ \` Assumption \\
  $\Psi \vd M \in \|g\ (Z_1\ \Psi^u)^1\ldots (Z_m\ \Psi^u)^1\|\hastype a$ for $g\not= h$
  \` Assumption \\
  $M = h \cdots$ \` By inversion on \texttt{grApp} \\
  $M = g \cdots$ \` By inversion on \texttt{grApp} \\
  $\perp$ \` Since $g\neq h$
 \end{tabbing}

 \item[Case] $\cD$ ends in $\mathtt{NotApp}_{2}^i$.
 \begin{tabbing}
  $\Psi \vd M \in \|h\ N_1^1\ldots N_n^1\|\hastype a$  \` Assumption \\
  $\Psi \vd M \in \|h\ (Z_1\ \Psi^u)^1\ldots (Z_{i-1}\ \Psi^u)^1 \ Q^{1}\ 
(Z_{i+1}\ \Psi^u)^1\ldots (Z_n\ \Psi^u)^1\| \hastype a$ and \\
  $\Psi \vd \mnot(N_i) \blip Q \hastype A_i$ \` Assumption \\
  $M = h\ M_1\ldots M_n$ and \\
  $\Psi \vd M_i \in \|N_i\| \hastype A_i$ \` By inversion \\
  $\Psi \vd M_i \in \|Q\| \hastype A_i$ \` By inversion \\
  $\perp$ \` By i.h.
 \end{tabbing}

 \item[Case] $\cD$ ends in $\mathtt{NotLam}$.
 \begin{tabbing}
 $\Psi\vd \mnot(\lxua N) \blip  \lxua Q\hastype A\TImp B$ \` This case \\
 $\Psi,x\oftp A\vd \mnot( N) \blip  Q\hastype B$ \` Subderivation \\
 $\Psi \vd \lxua M\in \| \lxua N\|\hastype A\TImp B$ \` Assumption \\
 $\Psi \vd \lxua M\in \| \lxua Q\|\hastype A\TImp B$ \` Assumption \\
 $\Psi,x\oftp A \vd M\in \| N\|\hastype B$\`  By inversion \\
 $\Psi,x\oftp A \vd M\in \| Q\|\hastype B$\`  By inversion \\
 $\perp$\` By i.h.
 \end{tabbing}
\end{description}
\end{proof}

Note that disjointness is based on exclusivity (Lemma~\ref{le:disj}), which
holds for \emph{any} strict term---it does not require simple terms.
Next, we turn to the other direction.  First a lemma concerning
the special case of generalized variables.

\begin{lemma}[Exhaustivity for Flexible Patterns]
\label{le:exhflex}
 \mbox{} \newline
 For every closed $M$ such that $\Psi \vd M \bua a$,
 either $\Psi \vd M\in \| E_A\ \Phi \| \hastype a$ or
 $\Psi \vd M\in \| Z\ \mnot_i(\Phi)\| \hastype a$ for some $i$.
\end{lemma}
\begin{proof}
  Assume $\Psi \vd M\bua a$.  Then by iterated applications of
  Lemma~\ref{le:exh} there exist $\OO$ and $\DD$ such that
  $\Psi=\OO,\DD$ and $\cdot;\OO;\DD\vd M\bua a$.
\begin{description}
\item[Case]
  For every $x\in\dom(\OO)$ we have $\Phi(x)\in\{0,u\}$
  and for every $x \in \dom(\DD)$ we have $\Phi(x)\in\{1,u\}$.
  Then $\Psi\vd M\in\| E\ \Phi\|$.
\item[Case]
  For some $x_i\in\dom(\OO)$ we have $\Phi(x_i) = 1$. \newline
  Then $\Psi \vd M\in\|Z\ x_{1}^{u}\ldots x_{i-1}^{u}\ 
  x_{i}^{1}\ x_{i+1}^{u}\ldots x_n^{u} \|$ and therefore $\Psi\vd
  M\in\|Z\ \mnot_i(\Phi)\|$.
\item[Case] For some $x_i \in\dom(\DD)$ we have $\Phi(x_i) = 0$. \newline
  Then $\Psi \vd M\in\|Z\ x_{1}^{u}\ldots x_{i-1}^{u}\ 
  x_{i}^{0}\ x_{i+1}^{u}\ldots x_n^{u} \|$ and therefore $\Psi\vd
  M\in\|Z\ \mnot_i(\Phi)\|$.
\end{description}
\end{proof}

We are now ready to prove exhaustivity of complementation.

\begin{lemma}[Exhaustivity of Complementation] \mbox{} \newline
\label{le:part2}
 Assume $\Psi\vd N\bua A$ is a simple linear pattern.  Then for every
closed $M$ such that $\Psi \vd M \bua A$, either $\Psi \vd M\in \| N \|
\hastype A$ or there is a $Q$ such that $\Psi\vd \mnot(N)\blip Q\hastype
A$ and $ \Psi \vd M\in \| Q\| \hastype A$.
\end{lemma}
\begin{proof}
  By induction on the structure of $\cD::\Psi \vd N\bua A$.
\begin{description}
 \item[Case] $\cD$ ends in \texttt{cPat}.  Then the claim
follows immediately by Lemma~\ref{le:exhflex}.

 \item[Case] $\cD$ ends in $\mathtt{c}\TImp I$. The i.h.\ yields the
two sub-cases.
 \begin{description}
  \item[Subcase] $\Psi,x\oftp A \vd M\in \| N\| \hastype B$.
  \begin{tabbing}
  $\Psi\vd\lxua M\in\|\lxua N\|\hastype A\TImp B$\` By rule \texttt{grLam}
  \end{tabbing}
  \item[Subcase] $\Psi,x\oftp A\vd\mnot(N)\blip Q\hastype B$ and 
  $\Psi,x\oftp A \vd M\in \| Q\| \hastype B$ for some $Q$.
  \begin{tabbing}
   $\Psi\vd\mnot(\lxua N)\blip\lxua Q \hastype A\TImp B$
   \` By rule $\mathtt{NotLam}$ \\
   $\Psi\vd\lxua M\in\|\lxua Q\|\hastype A\TImp B$ \` By rule \texttt{grLam}
  \end{tabbing}
\end{description}

 \item[Case] 
 \[ \cD=\ianc{\Psi \vd h\hastype A_1\RImp\cdots\RImp A_n\RImp a \quad
              \Psi \vd N_1\bua A_1 \quad \cdots \quad \Psi \vd N_n \bua A_n}
             {\Psi \vd h\ N_1^1\ldots N_{n}^1 \bua a}
             {\mathtt{c}{\RImp}E} \]
 First, assume $M = g\ M_1^1\ldots M_m^1$, for
$g\in\mathrm{dom}(\Sigma\cup\Psi)$, $h \not= g $. Then
\begin{tabbing}
 $\Psi\vd  \mnot (h\ N_1^1\ldots N_n^1) \blip  
 g\ (Z_{1} \ \Psi^{u})^{1}\ldots (Z_{m}\ \Psi^{u})^{1}\hastype a$
 \` By rule $\mathtt{NotApp}_1$ \\ 
 $\Psi\vd M_{i}\in\|Z_{i}\ \Psi^{u}\|\hastype A_i$ for all
$1\leq i\leq m$
 \` By rule \texttt{grFlx} \\ 
 $\Psi\vd g\ M_1^1\ldots M_m^1 \in \|g\ (Z_{1}\ \Psi^{u})^{1}\ldots
 (Z_{n}\ \Psi^{u})^{1}\|\hastype a$
 \` By rule \texttt{grApp}
\end{tabbing}
Otherwise, assume $M = h\ M_1^1\ldots M_n^1$. Again, the i.h.\ yields
two sub-cases.
\begin{description}
 \item[Subcase] $\Psi\vd M_i\in\|N_i\| \hastype A_i$,  for all  $1\leq i\leq n$.
 \begin{tabbing}
  $\Psi \vd h\ M_1^1\ldots M_n^1\in\|h\ N_1^1\ldots N_n^1\|\hastype
a$
 \` By rule \texttt{grApp}
\end{tabbing}
 \item[Subcase]
  $\Psi\vd\mnot(N_{i}) \blip  Q\hastype A_i$ and 
  $\Psi \vd M_{i}\in \| Q\| \hastype A_i$, for some $Q$.
 \begin{tabbing}
  $\Psi\vd M_{j}\in\|Z_{j}\ \Psi^{u}
  \|\hastype A_j$ for all  $j\not= i$, $1\leq j\leq n$
  \` By rule \texttt{grFlx} \\ 
  $\Psi\vd  \mnot (h\ M_1^1\ldots M_n^1)$ \\
  $\qquad \blip  
  h\ (Z_{1}\ \Psi^{u})^{1}\ldots (Z_{i-1}\ \Psi^{u} )^{1} \ Q^{1}\ 
  (Z_{i+1}\ \Psi^{u} )^{1}\ldots (Z_{n}\ \Psi^{u})^{1}\hastype a$ % \\
  \` By rule $\mathtt{NotApp}_2^i$ \\
  $\Psi \vd h\ M_1^1\ldots M_n^1\in \| h\ (Z_{1}\ \Psi^{u} )^{1}\ldots (Z_{i-1}\
\Psi^{u})^{1}\ Q^{1}\ (Z_{i+1}\ \Psi^{u} )^{1}\ldots (Z_{m}\ \Psi^{u}
)^{1} \| \hastype a$ \\
  \` By rule \texttt{grApp}.
\end{tabbing}
\end{description}
\end{description}
\end{proof}

The correctness of the algorithm for pattern complement follows
directly from the preceding two lemmas.

\begin{theorem}[Correctness of Pattern Complement]
\label{thm:compl}
\mbox{} \newline Assume $N$ is a simple linear pattern such that $\Psi
\vd N \hastype A$.  Then for every closed $M$ with $\Psi \vd M \bua A$,
$\Psi \vd M \in \|\mnot(N)\| \hastype A$ iff $\Psi \not\vd M \in
\|N\|$.
\end{theorem}

It is easy to see that simple linear patterns are closed under complementation.

\begin{theorem}[Closure under Complementation]
  Assume $M$ is a simple linear pattern with $\Psi \vd M \Uparrow A$.
  Then $\Psi\vd\mnot(M)\blip N\hastype A$ implies $N$ is a simple
  linear pattern and $\Psi \vd N \Uparrow A$.
\end{theorem}
\begin{proof}
  By induction on the structure of the derivation of
  $\Psi\vd\mnot(M)\blip N\hastype A$.
\end{proof}

\section{Unification of Simple Patterns}
\label{sec:sunif}

As we observed earlier, we can solve a relative complement problem by
pairing complementation with intersection.  We therefore address the
task of giving an algorithm for unification of linear simple patterns.
We start by determining when two labels are compatible:
 \begin{eqnarray*}
 1 \cap 1 = u \cap 1 = 1\cap u = 1\\ 
 0 \cap 0 = u \cap 0 = 0\cap u = 0\\ 
 u\cap u = u 
\end{eqnarray*}

 Recall that $\Phi$ is a list of labelled bound variables.  We call
$\Phi_1$ and $\Phi_2$ \emph{compatible} if they contain the same
variables in the same order, but with possibly different labels.  We can
extend the intersection operations to compatible lists.
 \begin{eqnarray*} 
\cdot\cap \cdot & = & \cdot\\
( \Phi, x^k)\cap ( \Phi', x^{k'}) & = & (\Phi\cap \Phi', x^{k\cap
k'}) \quad\mbox{\textrm{if $k\cap k'$ is defined.}} 
 \end{eqnarray*}

For contexts $\GG_1$ and $\GG_2$ that may have variable declarations
in common, we write $\GG_1 \cap \GG_2$ and $\GG_1 \cup \GG_2$ for
set-theoretic union and intersection.  In both cases we assume
that a variable $x$ declared in both $\GG_1$ and $\GG_2$ must be
assigned the same type in both contexts.

\begin{remark}
\label{rmk:compa}
Assume $\Phi_1$ and $\Phi_2$ are compatible and $\Phi_1 \cap \Phi_2$
is defined.  Then $\GG_1;\OO_1;\DD_1\vd\Phi_1\
ok$ and $\GG_2;\OO_2;\DD_2\vd\Phi_2\ ok$ implies
that $\DD_1 \cap \OO_2 = \DD_2 \cap \OO_1 = \emptyset$.
Moreover,
$(\GG_1\cap\GG_2);(\OO_1\cup\OO_2);(\DD_1\cup\DD_2)\vd (\Phi_1\cap\Phi_2) \ ok$.
From that it follows that $\Psi\vd M\in
\|E_{A}\ (\Phi_{1}\cap\Phi_{2})\| \hastype a$ iff
$(\GG_1\cap\GG_2);(\OO_1\cup \OO_2);(\DD_1 \cup \DD_2)\vd M\hastype a$.
\end{remark}

\begin{figure}
\[
\begin{array}{c}
\ianc{}
{\Psi\vd  (E_1 \ \Phi_1)\cap   (E_2\ \Phi_2)\blip  H\ (\Phi_{1}\cap \Phi_{2}) \hastype  a} 
{\cap \mathtt{FF}}
\\[2ex]
\mbox{\textrm{no rule for flex/flex same}}
\\[2ex]
\ianc{c\in\mathrm{dom}(\Sigma)\quad \Psi\vd  (H_1 \ \Phi_{1}) \cap 
M_{1}\blip N_{1}\hastype A_{1}\cdots \Psi\vd  (H_n \ \Phi_{n})\cap 
M_{n}\blip N_{n}\hastype A_{n}  } 
{ \Psi\vd  (E \ \Phi) \cap (c\ M_1^1\ldots M_n^1)\blip  c\ N_1^1\ldots N_n^1\hastype  a} {\cap
  \mathtt{FR}^c }
\\[2ex]
\ianc{y\in\mathrm{dom}(\Psi)\quad  \quad \Psi\vd  (H_1 \ \Phi_{1})\cap 
M_{1}\blip N_{1}\hastype A_{1}\cdots \Psi\vd  (H_n \ \Phi_{n})\cap 
M_{n}\blip N_{n}\hastype A_{n} } 
{\Psi\vd (E\ \Phi) \cap (y\ M_1^1\ldots M_n^1)\blip  y\ N_1^1\ldots N_n^1 \hastype  a} {\cap \mathtt{FR}^y }
\\[2ex]
\ianc{ h\in\mathrm{dom}(\Psi\cup\Sigma)\quad \Psi\vd  M_1 \cap N_{1} \blip
  Q_1\hastype  A_1 \cdots \Psi\vd  M_n \cap N_{n}\blip  Q_n \hastype  A_n} 
{\Psi\vd   (h\ M_1^1\ldots M_n^1)\cap (h\ N_1^1\ldots N_n^1) \blip  h\ 
Q_1^1\ldots Q_n^n \hastype  a}
{\cap \mathtt{RR}}
\\[2ex]
\ianc{\Psi,x\oftp A\vd   M  \cap  N \blip  Q\hastype  B}
{\Psi\vd  ( \lxua M ) \cap (\lxua) N \blip \lxua Q \hastype  A\TImp B}{\cap \mathtt{L}}
\end{array}
\]
\caption{Unification algorithm: $\Psi\vd M \cap N \blip Q \hastype A$}
  \label{fig:unif}
\end{figure}

\begin{definition}[Higher-Order Pattern Intersection]
 Assume $M$ and $N$ are linear simple patterns without shared existential
variables such that $\Psi \vd M \Uparrow A$ and $\Psi \vd N\Uparrow A$.  We
define $\Psi\vd M \cap N \blip Q \hastype A$ by the rules in
Figure~\ref{fig:unif}, where the $H$'s are fresh variables of appropriate
type. We omit two rules, $\cap \mathtt{RF}^c$ and $\cap \mathtt{RF}^y$, that
are symmetric to $\cap \mathtt{FR}^c$ and $\cap \mathtt{FR}^y$.

The rules $\cap \mathtt{FR}^c$ and $\cap\mathtt{RF}^c$ have the
following proviso: for all $1 \leq i \leq n$, $\dom(\Phi_i) =
\dom(\Phi)$ and
\begin{eqnarray*}
\forall x. \Phi(x) = 0  \Imp  \forall i\ldot \Phi_{i}(x) = 0\\ 
\forall x. \Phi(x) = u  \Imp  \forall i\ldot \Phi_{i}(x) = u\\ 
\forall x. \Phi(x) = 1  \Imp  \exists i\ldot \Phi_{i}(x) = 1
\And \forall j\ldot j\not = i \Imp \Phi_{j}(x) = u
\end{eqnarray*}
The rules $\cap \mathtt{FR}^y$ and $\cap RF^y$ are subject to the
proviso:
\begin{eqnarray*}
\forall x. \Phi(x) = 0  \Imp  \forall i\ldot \Phi_{i}(x) = 0\\ 
\forall x. \Phi(x) = u  \Imp  \forall i\ldot \Phi_{i}(x) = u\\ 
\forall x. x \not= y \And \Phi(x) = 1  \Imp  \exists i\ldot \Phi_{i}(x) = 1
\And \forall j\ldot j\not = i \Imp \Phi_{j}(x) = u \\
\Phi(y)=1 \Imp  \forall i. \Phi_{i}(y) = u
\end{eqnarray*}
Finally define $\Psi \vd M \cap N = {\cal Q} \hastype A$ if ${\cal Q}=
 \{Q\mid\Psi\vd  M\cap N \blip  Q \hastype  A\}$.
\end{definition} 

Some remarks are in order:
\begin{itemize}
 \item In rule $\cap \mathtt{FF}$ we can assume $\Phi_1$ and $\Phi_2$
are compatible lists of variables, since generalized variables
are fully applied and their arguments are in a standard order.
 \item Since patterns are linear and $M$ and $N$ share no pattern
variables, the flex/flex case arises only with distinct variables.  This
also means we do not have to apply substitutions or perform the
customary occurs-check.
 \item In the flex/rigid and rigid/flex rules, the proviso enforces the
typing discipline since each strict variable $x$ must be strict in some
premise.  If instead $y$ is the projected variable, the modified
condition on $y$ takes into account that the head of an application
constitutes a strict occurrence; moreover, since $y$ did occur, it is
set to $u$ in the rest of the computation, as there are no more
requirements on that variable.
 \item The symmetric rules take the place of an explicit exchange rule
that is problematic with respect to termination. 
\end{itemize}

The following example illustrates how the flex/rigid rules, in this 
case $\cap \mathtt{FR}^c$, make unification on simple patterns finitary.
We describe a \emph{unification problem} by omitting
the eventually computed solution as $\Psi \vd M \cap N \hastype A$.

\begin{example}
 Consider the unification problem
  \[ x\oftp a\vd E\ x^{1} \cap c\ (F\ x^{u})^{1} \ (F'\ x^{u})^{1}\hastype a \]
 Since $x$ is strict in the left-hand side, there are two ways to
apply the $\cap \mathtt{FR}^c$ rule, leading to the following subproblems:
\[
\begin{array}{rlrr}
1. &x\oftp a\vd E' \ x^{1}\cap\  F\  x^{u} \hastype a 
& & x\oftp a\vd E''\ x^{u}\cap\  F'\  x^{u} \hastype a
\\
2. & x\oftp a\vd E' \ x^{u}\cap\  F\  x^{u} \hastype a
& & x\oftp a\vd E''\ x^{1}\cap\  F'\  x^{u} \hastype a
\end{array}
\]
Hence the result:
\begin{eqnarray*}
%\label{eq:int}
x\oftp A\vd E\ x^{1} \cap c\ (F\  x^{u})^{1} \ (F'\  
x^{u})^{1}=  \{c\ (H\  x^{1})^{1} \ (H'\  x^{u})^1,c\ (H\  x^{u})^1 \ (H'\  
x^{1})^1\}
\end{eqnarray*}
\end{example}

Note that, similarly to complementation, intersection returns
a set of patterns with common terms; again it is possible, in a
post-processing phase to make the result exclusive.

 The following example illustrates the additional proviso on
$\cap \mathtt{FR}^{y}$
\begin{example}
  The unification problem 
  \[ y\oftp a \RImp a \RImp a \vd E\ y^{0} \cap y\ (F\ 
  y^{1})^1 \ (F'\ y^{u})^1 \hastype a \]
  has no solution, whereas
 \[ y\oftp a \RImp a \RImp a
  \vd E\ y^1 \cap y\ (F\ y^1)^1 \ (F'\ y^{0})^1 = \{y \ (H\ y^1)^1 \ 
  (H'\ y^{0})^1\} \hastype a \]
\end{example}

This first lemma will be needed to handle the case for unification
of generalized variables.

\begin{lemma}
\label{le:flexunif}
Assume $\Phi_1$ and $\Phi_2$ are compatible and $\Phi_1 \cap \Phi_2$
is defined.  Assume furthermore that $\GG_1;\OO_1;\DD_1\vd\Phi_1\
ok$ and $\GG_2;\OO_2;\DD_2\vd\Phi_2\ ok$.
Then $\GG_1; \OO_1; \DD_1 \vd M\hastype A$ and $\GG_2; \OO_2; \DD_2 \vd M\hastype A$ iff $(\GG_{1}\cap\GG_{2});(\OO_{1}\cup\OO_{2});(\DD_{1}\cup\DD_{2}) \vd
M\hastype A$.
\end{lemma}
\begin{proof}
  From left to right by induction on the size of $(\GG_1\cup\GG_2)\setminus
(\GG_1\cap\GG_2)$, using tightening (Lemma~\ref{le:exh}).  From right to
left by appropriate appeals to loosening (Lemma~\ref{thm:loosex}).
\end{proof}

 We introduce two $n$-ary strict application rules which, for the
special case of simple terms, capture the notion of atomic forms more
compactly than the previous definition.  The rules differ only in
whether the head $h$ of the atomic term is a strict variable or
unrestricted.  These will be needed in the proof of Lemma~\ref{le:unif2}
and Lemma~\ref{le:unif1}.

\begin{eqnarray*}
 \ianc{(\GG,\DD^{u}_{i});\OO;\DD_i^1 \vd M_{i}\hastype A_{i} \quad 
       1\leq i\leq n}
      {\GG;\OO;\DD \vd h\ M_1^1\ldots M_{n}^1\hastype B}
      {{\RImp}E^{u}}
\end{eqnarray*}
where $h\ \hastype A_1\RImp\ldots\RImp A_{n}\RImp b$ in $\dom(\GG\cup\Sigma)$
and
\begin{enumerate}
 \item $\forall x\in\dom(\DD)\ldot \exists\mbox{!} i:1\leq i\leq n \ldot
  x\in\dom(\DD_{i}^1)$.
 \item $\forall i:1\leq i\leq n\ldot (\DD_{i}^{u},\DD_{i}^1)=\DD$.
\end{enumerate}

%proj rule
\begin{eqnarray*} 
\ianc{(\GG,\DD^{u}_{i});\OO;\DD_i^1 \vd M_{i}\hastype A_{i}\quad 
1\leq i\leq n}{\GG;\OO; \DD\vd y\ M_1^1\ldots M_{n}^1\hastype 
B}{{\RImp}E^{1}} 
\end{eqnarray*}
where $y \hastype A_1\RImp\ldots\RImp A_{n}\RImp b \in \dom(\DD)$
and
\begin{enumerate}
 \item $\forall x\in\dom(\DD), x\not= y\ldot\exists\mbox{!} i:1\leq i\leq n
\ldot x\in\dom(\DD_{i}^1)$.
 \item $\forall i:1\leq i\leq n \ldot(\DD_{i}^{u},\DD_{i}^1)=\DD$. 
 \item $\forall i:1\leq i\leq n \ldot y\in\dom(\DD_i^u)$.
\end{enumerate}

It is straightforward, but tedious to show that these rules can replace
the rules for atomic terms.  The curious reader is invited to
consult~\cite{Momigliano00phd} for details.

We are now ready to address correctness of unification.
First we show that our algorithm only computes unifiers, then that
the set of unifiers we compute is most general.

\begin{lemma}[Intersection Computes Unifiers]
\label{le:unif2}
\mbox \newline
For any simple linear pattern $N_1$ and $N_2$ without shared variables
such that $\Psi \vd N_1  \Uparrow A$ and $\Psi \vd N_2 \Uparrow A$, for
every $N$ such that $\Psi\vd N_1\cap N_{2}\blip N$ if $\Psi\vd
M\in\|N\|\hastype A$, then $ \Psi\vd M\in\|N_1\|\hastype A$ and
$\Psi\vd M\in\|N_{2}\|\hastype A$.
\end{lemma}
\begin{proof}
By induction on the structure of $\cD::\Psi\vd N_1\cap N_{2}\blip  N$
and inversion on $\cD'::\Psi\vd M\in\|N\|\hastype A$.  We show only
some of the cases; the others are analogous.
\begin{description}
 \item[Case] $\cD$ ends in $\cap \mathtt{FF}$:
 \begin{tabbing}
 ${\Psi\vd  (E_1 \ \Phi_1)\cap   (E_2\ \Phi_2)\blip  H\ (\Phi_{1}\cap
  \Phi_{2}) \hastype  a} $ \` Assumption \\
 $\Psi\vd M\in \| H\ (\Phi_{1}\cap \Phi_{2})\| \hastype  a$
 \` Assumption \\
 $\GG_i;\OO_i;\DD_i\vd\Phi_i\ ok$ for $i=1,2$ for some $\GG_i$, $\OO_i$,
$\DD_i$
  \` Determined from $\Phi_i$ \\
 $(\GG_{1}\cap\GG_{2});(\OO_{1}\cup\OO_{2});(\DD_{1}\cup\DD_{2})
\vd\Phi_1\cap\Phi_2\ ok$\` By Remark~\ref{rmk:compa} \\
 $\GG_i;\OO_i;\DD_i\vd M \hastype a$
 \` By  Lemma~\ref{le:flexunif} ($\leftarrow$)\\
 $\Psi\vd M\in\|N_i\|\hastype a$ \` By rule \texttt{grFlx}
\end{tabbing}
\item[Case] $\cD$ ends in  $\cap \mathtt{FR}^c $.
\begin{tabbing}
 $\cD::\Psi\vd (E\ \Phi)\cap (c\ Q_1^1\ldots Q_{n}^1)
\blip  c\ N_1^1\ldots N_n^1\hastype a$
 \` Assumption \\
 $\cD_i::\Psi\vd (E\ \Phi_i)\cap Q_i \blip N_i\hastype A_i,$ for all $ 1\leq i\leq n$
 \` Subderivations \\
 $\Psi\vd c\ M_1^1\ldots M_n^1\in\|c\ N_1^1\ldots N_n^1\|\hastype a$ \` Assumption \\
 $\Psi\vd M_i\in\|N_i\|\hastype A_i$ \` By inversion \\
 $\Psi\vd M_{i}\in\|Q_{i}\|\hastype A_i$ and $\Psi\vd M_{i}\in\|E_{i}\
\Phi_{i}\|\hastype A_i$
 \` By i.h.\ on $\cD_i$ \\
 $(\GG_{i},\DD^u_i);\OO;\DD^1_i\vd \Phi_i\ ok$ and $(\GG_{i},\DD^u_i);\OO;\DD^1_i\vd M_{i}\hastype A_{i}$
 \` By rule \texttt{grFlx} \\
 $\god\vd c\ M_1^1\ldots M_n^1\hastype a$ \` By  rule ${\RImp}E^{u}$ \\
 $\Psi\vd c\ M_1^1\ldots M_n^1\in\|E\ \Phi\|\hastype a$
 \` By rule \texttt{grFlx} \\
 $\Psi\vd c\ M_1^1\ldots M_n^1\in\|c\ Q_1^1\ldots Q_n^1\| \hastype a$
 \` By rule \texttt{grApp}
\end{tabbing}
\end{description}
\end{proof}

The second part consists of showing that any unifier of two patterns
is an instance of an element from the computed set of unifiers.

\begin{lemma}[Intersections are Most General]
\label{le:unif1}
 For any simple linear patterns $N_1$ and $N_2$ without shared variables such
that $\Psi \vd N_1 \bua A$ and $\Psi \vd N_2 \bua A$, if $\Psi\vd
M\in\|N_1\|\hastype A$ and $\Psi\vd M\in\|N_{2}\|\hastype A$, then there is
$N$ such that $\Psi\vd N_1\cap N_{2}\blip N\hastype A$ and $\Psi\vd
M\in\|N\|\hastype A$.
\end{lemma}
\begin{proof}
 By simultaneous induction on the structure
  of $\cD_1::\Psi \vd M\in\|N_1\| \hastype A$ and $\cD_2::\Psi \vd
  M\in\| N_2\| \hastype A$.
\begin{description}
\item[Case] $\cD_1,\cD_2$ end in \texttt{grFlx}:
\begin{tabbing}
 $\GG_i;\OO_i;\DD_i\vd\Phi_i\ ok$ and $\GG_i;\OO_i;\DD_i\vd M \hastype a$
 for $i=1,2$ \` Subderivations \\
 $\Phi_1 \cap \Phi_2$ is defined \` By exclusivity (Lemma~\ref{le:disj}) \\
 $\Psi\vd  (E_1 \ \Phi_1)\cap   (E_2\ \Phi_2)\blip  H\ (\Phi_{1}\cap
\Phi_{2}) \hastype  a$ \` By rule $\cap\mathtt{FF}$ \\
 $(\GG_{1}\cap\GG_{2});(\OO_{1}\cup\OO_{2});(\DD_{1}\cup\DD_{2})\vd M\hastype a$\` By Lemma~\ref{le:flexunif}($\arrow$)\\ 
 $(\GG_{1}\cap\GG_{2});(\OO_{1}\cup\OO_{2});(\DD_{1}\cup\DD_{2})\vd\Phi_1\cap\Phi_2\ ok$\` By Remark~\ref{rmk:compa} \\
 % $\Psi\vd M\in\|E_i\ \Phi_i\|\hastype a$ for $1\leq i\leq 2$\bh
 $\Psi\vd M\in \| H\ (\Phi_{1}\cap \Phi_{2})\| \hastype  a$
 \` By rule \texttt{grFlx}
\end{tabbing}
\item[Case]
 $\cD_1$ ends in \texttt{grFlx} and $\cD_2$ ends in \texttt{grApp}: there are
two cases depending on whether the head of $N_2$ is a constant or a parameter.
\begin{description}
\item[Subcase] The head of $N_2$ is a constant $c$.
\begin{tabbing}
 $\Psi\vd M \in \|c\ Q_1^1\ldots Q_n^1\| \hastype a$ \` Assumption \\
 $M = c\ M_1^1\ldots M_n^1$ and $\cD^2_i::\Psi\vd M_{i}\in\|Q_{i}\|\hastype A_i$ for
 all $1\leq i\leq n$ \` Subderivation \\
 $\Psi\vd c\ M_1^1\ldots M_n^1\in\|E\ \Phi\|\hastype a$ \` Assumption \\
 $\god\vd c\ M_1^1\ldots M_n^1 \hastype a$ and  $\god\vd\Phi\ ok$
 \` By inversion on rule \texttt{grFlx}\\ 
 $(\GG,\DD^u_i);\OO;\DD^1_i\vd M_{i}\hastype A_{i}$ for some
 $\DD^u_i,\DD^u_i$ satisfying $(1)$ and $(2)$ \\
 \` By inversion on rule ${\RImp}E^u$ \\ 
 $\cD^1_i::\Psi\vd M_{i}\in\|E_{i}\ \Phi_{i}\|\hastype A_i$
 for  $\Phi_i$ such that $(\GG,\DD^u_i);\OO;\DD^1_i\vd\Phi_i \ ok$\\
 \` By rule \texttt{grFlx} \\
 $\cD_i::\Psi\vd (E_{i}\ \Phi_i)\cap Q_i \blip  N_i\hastype A_i$ and
 $\Psi\vd M_{i}\in\|N_{i}\|\hastype A_i$ 
 \` By i.h.\ on $\cD_i^1,\cD_i^2$ \\
 $\cD::\Psi\vd (E\ \Phi)\cap (c\ Q_1^1\ldots Q_n^1) \blip c\
N_1^1\ldots N_n^1 \hastype a$
 \` By rule $\cap \mathtt{FR}^c$\\
 $\Psi\vd c\ M_1^1\ldots M_n^1\in\|c\ N_1^1\ldots N_n^n\|\hastype a$
 \` By rule \texttt{grApp}
\end{tabbing}
 \item[Subcase]  Proceed as above, but using inversion
 on rule ${\RImp}E^{1}$ % \ie Corollary~\ref{co:derivey}.
\end{description}
\item[Case] $\cD_2$ ends in  \texttt{grFlx} and $\cD_1$ ends in \texttt{grApp}:
symmetrical to the above.
\item[Case] $\cD_1,\cD_2$ end in \texttt{grLam}: straightforward
by induction hypothesis.
\item[Case] $\cD_1,\cD_2$ end in \texttt{grApp}: a straightforward appeal to 
the induction hypothesis as in the above case.
\end{description}
\end{proof}

The correctness of the algorithm for pattern intersection follows
directly from the preceding two lemmas.

\begin{theorem}[Correctness of Pattern Intersection]
\label{thm:unif} \mbox{} \newline
 Assume $N_1$ and $N_2$ are simple linear patterns without shared variables
such that $\Psi \vd N_1 \Uparrow A$ and $\Psi \vd N_2 \Uparrow A$.  Then $\Psi\vd
M\in\|N_1\|\hastype A$ and $\Psi\vd M\in\|N_{2}\|\hastype A$ iff $\Psi\vd
M\in\|N_1 \cap N_2\|\hastype A$.
\end{theorem}

Also note that the intersection of linear simple patterns is again
a simple linear pattern.

\begin{theorem}[Closure under Intersection]
 Assume $M$ and $N$ are simple linear patterns with $\Psi \vd
M \Uparrow A$ and $\Psi \vd N \Uparrow A$.  Then $\Psi \vd M \cap N \blip
Q \hastype A$ implies that $Q$ is a simple linear pattern
and $\Psi \vd Q \Uparrow A$.
\end{theorem}
 \begin{proof}
  By induction on the structure of the derivation of $\Psi \vd M \cap N
\blip Q \hastype A$.
 \end{proof}

\section{The Algebra of Linear Simple Patterns}
\label{sec:algebra}

An interesting and natural question is whether complementation is
involutive.  The answer is of course positive, since the latter is a
boolean property and the complement operation has been shown to satisfy
``tertium non datur'' and the principle of non-contradiction.  However,
the reader should keep in mind that the \emph{representation} of the set
$\mnot(\mnot(N))$ may be different from $\{N\}$, even though the two
sets are guaranteed to have the same set of ground instances.  Since on
finite set of patterns we also have intersection and set-theoretic
union, we obtain a boolean algebra.  For the sake of readability, we
introduce the following notation: $\Pat$ denotes the finite set of
linear simple patterns $M$ with $\Psi \vd M \hastype A$.  In the
following, we also drop the type information and overload the singleton
pattern notation.

\begin{definition}
\label{def:setop} For
$\cM,\cN \in \Pat$, define: 
\begin{eqnarray*}      
 \cM \cap \cN & = & \bigcup_{M\in\cM,N\in\cN} M \cap N \\[2ex]
 \mnot (\cM) & = & \bigcap_{M\in\cM} \mnot(M)
\end{eqnarray*}
\end{definition}

Those operations on sets of patterns satisfy the same properties that
singleton intersection and complementation do.

\begin{corollary}[Correctness of Set Intersection] \mbox{} \newline
\label{co:setunif}
For $\cN_1,\cN_{2}\in \Pat$, $\Psi\vd M\in\|\cN_1\|\hastype A$ and $\Psi\vd
M\in\|\cN_{2}\|\hastype A$ iff $\Psi\vd M\in\|\cN_1\cap
\cN_{2}\|\hastype A$.
\end{corollary}
% \begin{proof}
%   $\GG\vd M\in\|\cN_1\|\hastype A$ and $\GG\vd
%   M\in\|\cN_{2}\|\hastype A$ iff there is $N_1\in\cN_1$ and
%   $N_{2}\in\cN_{2}$ such that $\GG\vd M\in\|N_1\|\hastype A$ and
%   $\GG\vd M\in\|N_{2}\|\hastype A$ iff, by Corollary \ref{co:unif},
%   $\GG\vd M\in\|N_1\cap N_{2}\|\hastype A$ iff, by definition,
%   $\GG\vd M\in\|\cN_1\cap \cN_{2}\|\hastype A$.
% \end{proof}

\begin{corollary}[Correctness of Set Complement] \mbox{} \newline
\label{co:setpart}
For $\cN \in \Pat$, $\Psi \vd M \in \|\mnot(\cN)\| \hastype A$ iff
$\Psi \not\vd M \in \|\cN\| \hastype A$
\end{corollary}

As we have remarked earlier, we can define the relative complement
operation by using complement and intersection.  Its correctness
follows immediately from the correctness of pattern set intersection
and complement.

\begin{definition}[Relative Complement]\mbox{}\\
\label{def:relcompl}
Given $\cM,\cN\in\Pat$, we define \mbox{$\cM - \cN = \cM\cap \mnot (\cN)$}.
\end{definition}

The properties above mean that we can organize, for a given signature
$\Sigma$, context $\Psi$, and a type $A$, finite sets of simple linear
patters into a Boolean algebra by taking equality as extensional
identity on sets of terms without existential variables.  In symbols, for
$\cN_1,\cN_{2}\in \Pat$:
\begin{eqnarray*}
 \cN_1\simeq \cN_2 & \mbox{iff} & \|\cN_1\| = \|\cN_2\|
\end{eqnarray*}

Under this interpretation, the $\zero$ element is the empty set and the
$\one$ element the singleton set containing the $\eta$-expansion of a
generalized existential variable of the appropriate type that may
depend on all variables in the context $\Psi$.
\begin{eqnarray*}
 \zero & = & \emptyset \\
 \one & = & \{\lam x_1^u\oftp A_1\ldots \lam x_n^u\oftp A_n\ldot E\ \Psi^u\
x_1^u\ldots x_n^u\}
\end{eqnarray*}
 where $A = A_1\TImp \cdots A_n \TImp a$.

\begin{theorem}
\label{thm:algebra}
 Consider the algebra $\langle \Pat,\cup,\cap,\mnot, \one,
\zero\rangle$. Then the following holds:
\begin{enumerate}
 \item $\cM\cap\cM \simeq \cM$.
 \item $\cM\cap\cN \simeq \cN\cap\cM$.
 \item\label{it:DM} $\cM\cap(\cN\cup\cP)\simeq
   (\cM\cap\cN)\cup(\cM\cap\cP)$.
 \item $\cM\cap(\cN\cap\cP) \simeq (\cM\cap\cN)\cap\cP$.
 \item $\mnot(\mnot (\cM))\simeq\cM$.
 \item $\mnot(\one)\simeq \zero$.
 \item $\mnot(\zero)\simeq \one$.  
\end{enumerate}
\end{theorem}
\begin{proof}
  From Corollaries~\ref{co:setunif} and~\ref{co:setpart} and the fact that
  $\cup$ is set-theoretic.
\end{proof}

\begin{corollary}
\label{co:boolean}
 The algebra of finite sets of simple linear patterns is boolean.
\end{corollary}

It is notable that the $\cup$ operator must be set-theoretic union
rather than anti-unification or generalization, as traditional in
lattice-theoretic investigations of the algebra of terms
\cite{Lassez88}. The problem is the intrinsically classical
nature of complementation which is not compatible with the very
irregular structure of the lattice of terms where the smallest upper
bound is interpreted as anti-unification.

  We end this section showing how pattern complement can be used as a
building block of our main application, that is a clause complement
algorithm~\cite{Bar90}. In (higher-order) logic programming, in fact,
pattern complement is a necessary component in any algorithm to
synthesize the negation of a given program.  This synthesis includes two
basic operations: negation to compute the complements of heads of
clauses in the definition of a predicate, and intersection to combine
results of negating individual clause heads.  In this paper we have
provided algorithms to compute both. A full development for the
higher-order case can be found in~\cite{Momigliano00phd}.

\begin{example}
\label{ex:combi}
We can combine Example~\ref{ex:lam} and~\ref{ex:etardx} and consider
the following trivial program, which encodes when an object-level
lambda term is a $\beta\eta$-redex:
\[
\begin{array}{lll}
 {\it betardx} & : & {\it isredx}\ (\itapp\ (\itlam\ (\lamb{x^u}{\itexp} E\ x^u))\ F).\\
 {\it etardx} & : & {\it isredx}\ (\itlam\ (\lamb{x^u}{\itexp} \itapp\ (E\ x^0) \ x)).
\end{array}\]
We can compute the complement of both heads, as follows:
\[
\begin{array}{ll}
 \multicolumn{2}{l}{\mnot\{\itapp\ (\itlam\ (\lamb{x^u}{\itexp} E\ x^u)) \ 
 F, \itlam(\lamb{x^u}{\itexp} \itapp\ (E\ x^0) \ x)\}} \\
 = & \mnot(\itapp\ (\itlam\ (\lamb{x^u}{\itexp} E\ x^u))\ F)
 \cap\mnot(\itlam(\lamb{x^{u}}{\itexp} \itapp\ (E\ x^{0})\ x )) \\
 = & \{ \itlam\ (\lamb{x^u}{\itexp} H\ x^u), \itapp\ (\itapp\ H\ H') \ H''\} \\
   & \null \cap \{
  \begin{array}[t]{@{}l}
   \itlam\ (\lamb{x^u}{\itexp} \itapp\ (H\ x^1)\ (H'\ x^u)),\\
   \itlam\ (\lamb{x^u}{\itexp} \itapp\ (H\ x^u)\ (\itapp\ (H'\ x^u) \ (H'' x^u))), \\
   \itlam\ (\lamb{x^u}{\itexp} \itapp\ (H\ x^u)\ (\itlam\ (\lamb{y^u}{\itexp} H'\ x^u\ y^u))),\\
   \itlam\ (\lamb{x^u}{\itexp} \itlam\ (\lamb{y^u}{\itexp} H\ x^u\ y^u) ),\\
   \itlam\ (\lamb{x^u}{\itexp} x),\\
   \itapp\ H\ H' \}
  \end{array} \\
 = & \{
   \begin{array}[t]{@{}l}
    \itlam\ (\lamb{x^u}{\itexp} \itapp\ (H\ x^1)\ (H'\ x^{u})), \\ 
    \itlam\ (\lamb{x^u}{\itexp} \itapp\ (H\ x^u)\ (\itapp\ (H'\ x^{u}) \ ( H''\ x^{u}))),\\
    \itlam\ (\lamb{x^u}{\itexp} \itapp\ (H\ x^u)\ (\itlam\ (\lamb{y^u}{\itexp} H'\ x^u\ y^u)))\\
    \itlam\ (\lamb{x^u}{\itexp} \itlam\ (\lamb{y^u}{\itexp} H'\ x^u\ y^u)),\\
    \itlam\ (\lamb{x^u}{\itexp}x),\\
    \itapp\ (\itapp\ H\ H')\  H''\}
   \end{array}
\end{array}
\]

This yields the negation of that program, that is the complementary clauses:
\[
\begin{array}{lll}
 {\it nb}_1 & : & {\it non\_isredx}\ (\itlam\ (\lamb{x^u}{\itexp} \itapp\ (H\
x^1)\ (H'\ x^{u}))).\\
 {\it nb}_2 & : & {\it non\_isredx}\ (\itlam\ (\lamb{x^u}{\itexp} \itapp\ (H\
x^u)\
 (\itapp\ (H'\ x^{u}) \ ( H''\ x^{u})))).\\
 {\it nb}_3 & : & {\it non\_isredx}\ (\itlam\ (\lamb{x^u}{\itexp} \itapp\ (H\
x^u)\ (\itlam\ (\lamb{y^u}{\itexp} H'\ x^u\ y^u)))).\\
 {\it nb}_4 & : & {\it non\_isredx}\ (\itlam\ (\lamb{x^u}{\itexp}
\itlam\ (\lamb{y^u}{\itexp} H\ x^u\ y^u) )).\\
 {\it nb}_5 & : & {\it non\_isredx}\ (\itlam\ (\lamb{x^u}{\itexp}x)).\\ 
 {\it nb}_6 & : & {\it non\_isredx}\ (\itapp\ (\itapp\ H\ H')\ H'').
\end{array}\]
\end{example}

\section{Conclusions}
\label{sec:concl}
In this paper we have been concerned with the relative complement
problem for higher-order patterns.  As we have seen, the complement
operation does not generalize easily from the first-order case.
Indeed, the complement of a partially applied higher-order pattern
cannot be described by a pattern, or even a by finite set of patterns.
The formulation of the problem suggests that we should consider a
$\lambda$-calculus with an internal notion of \emph{strictness} so that
we can directly express that a term must depend on a given variable. We
have developed such a calculus and we have shown that via a suitable
embedding in our calculus the complement of a linear pattern is a finite
set of linear patterns and unification of two patterns is decidable and
leads to a finite set of most general unifiers.  Moreover, they form a
boolean algebra under set-theoretic union, intersection (implemented via
unification) and the complement operation.

The latter item brings up the question if we can actually decide
extensional equality between, and membership of terms in, finite sets of
simple terms.  For membership, one can see that $\Psi\vd M\in \|
N_1,\ldots,N_n\|$ iff $M$ unifies with some $N_i$.  As far as equality
is concerned between say $\| M_1,\ldots,M_m\|$ and $\| N_1,\ldots,N_n\|$
calculate the two relative complements $\{ M_1,\ldots,M_m\} - \{
N_1,\ldots,N_n\}$ and $ \{ N_1,\ldots,N_n\} - \{ M_1,\ldots,M_m\}$ and
then check if they are both empty.  An emptiness check would rely on the
decidability of inhabitation in the underlying calculus.  We conjecture
this question to be decidable for the strict $\lambda$-calculus and we plan to
address this question in future work.
 % possibly via a translation in an equivalent
 % contraction-free sequent calculus, following Dyckhoff's
 % approach~\cite{Dyckhoff92}.

Our main application is the transformational approach to negation in
higher-order logic programming \cite{Bar90}, where pattern complement
and unification is a necessary component. We plan to extend the
results to dependent types to endow intentionally weak frameworks such
as Twelf \cite{Schurmann98cade} with a logically meaningful notion of
negation along these lines.

It may be argued that the restriction to simple terms is somewhat ad
hoc.  Ideally, one would have a complement algorithm for the full strict
lambda-calculus (including vacuous types). Yet, this seems to be
ill-defined, because ``occurrence'' no longer has the desired meaning
once we lift the principle that constructors should be strict in their
argument.  As we have remarked earlier, it is possible to describe
complement and unification algorithms for a larger fragment than treated
here by allowing arbitrary abstractions, if we adhere to the above
strictness assumption for constructors.  The technical development is
not difficult but entails a proliferation of rules to cover the new
abstraction cases, as well as the duplication of all rules concerning
strict application in versions similar to the ${\RImp}E^u$ and
${\RImp}E^1$ typing rules.

Finally, it is our contention that the strict $\lam$-calculus that we
have introduced has independent interest in the investigation of
sub-structural logics.  Our type system is simple and uniform and
arguably more elegant than those ones presented in the literature (see
the earlier discussion of related work at the end of
Section~\ref{sec:canonthm}).  Moreover, the explicit introduction of the
notion of \emph{vacuous} or \emph{irrelevant} variables can be useful in
a variety of contexts.  In fact, the second author has suggested some
unexpected usage of those variables in type theory for uses in reasoning
about staged computation~\cite{Pfenning00saig} and proof compression in
logical frameworks~\cite{Pfenning01lics}.  Furthermore, extending a
linear $\lam$-calculus with vacuous variables permits more programs
under type assignment; for example a term such as $\lam x\ldot \lam
y\ldot x\otimes (\lam w\ldot y)\ x$, which is traditionally considered
\emph{not} linear, can be given the linear type $A \lop B \lop (A
\otimes B)$.  This carries over to the study of \emph{explicit
substitutions} in resource-conscious $\lam$-calculi~\cite{Ghani98} where
it might clarify the logical status of the \emph{extension operator}.

\begin{acknowledgments}
  We would like to thank Roberto Virga, who discovered an error in an
  earlier version of this paper, and Iliano Cervesato and Carsten
  Sch\"urmann for several discussions and comments on a draft of this
  paper.
\end{acknowledgments}

\bibliographystyle{acmtrans}
% \bibliography{not,framework,strict}
\bibliography{tocl}

\begin{received}
Submitted September 2001
% November 1993;
% accepted January 1996
\end{received}

\end{document}